\documentclass[aip,aplphotonics,preprint, citeautoscript, showpacs,floatfix,superscriptaddress]{revtex4-1}
\usepackage{bm}
\usepackage{amssymb}
\usepackage{graphicx}
\usepackage{amsmath}
\usepackage{textcomp}
\usepackage{multirow}
\usepackage{ulem}

\newcommand{\Red}[1]{#1}

\begin{document}

\title{Perspective: Intraband divergences in third order optical \Red{response} of 2D systems}

\author{J. L. Cheng}
\email{jlcheng@ciomp.ac.cn}
\affiliation{The Guo China-US Photonics Laboratory, Changchun Institute of Optics,
fine Mechanics and Physics, Chinese Academy of Sciences, 3888 Eastern
South Lake Road, Changchun, Jilin 130033, China.}

\affiliation{University of Chinese Academy of Sciences, Beijing 100049, China}

\author{J. E. Sipe}
\email{sipe@physics.utoronto.ca}
\affiliation{Department of Physics and Institute for Optical Sciences, University of Toronto, 60 St. George
Street, Toronto, Ontario M5S 1A7, Canada}

\author{S. W. Wu}
\email{swwu@fudan.edu.cn}
\affiliation{State Key Laboratory of Surface Physics, Key Laboratory of Micro
and Nano Photonic Structures (MOE), and Department of Physics, Fudan
University, Shanghai 200433, China}

\author{Chunlei Guo}
\email{chunlei.guo@rochester.edu}
\affiliation{The Guo China-US Photonics Laboratory, Changchun Institute of Optics,
fine Mechanics and Physics, Chinese Academy of Sciences, 3888 Eastern
South Lake Road, Changchun, Jilin 130033, China.}

\affiliation{The Institute of Optics, University of Rochester, Rochester, NY 14627,
USA.}


\begin{abstract}
\Red{The existence of large nonlinear optical coefficients is one of
the preconditions for using nonlinear optical materials in nonlinear
optical devices. For a crystal, such large coefficients can be achieved
by matching photon energies with resonant energies between different
bands, and so the details of the crystal band structure play an important
role. Here we demonstrate that large third-order nonlinearities can
also be generally obtained by a different strategy: As any of the
incident frequencies or the sum of any two or three frequencies approaches
zero, the doped or excited populations of electronic states lead to
divergent contributions in the induced current density. We refer to
these as intraband divergences, by analogy with the behavior of Drude
conductivity in linear response.} \Red{Physically, such resonant processes}
can be associated with a combination of inraband and interband optical
transitions. Current-induced second order nonlinearity, coherent current
injection, and jerk currents are all related to such divergences,
\Red{and} we find similar divergences in degenerate four wave mixing and
cross-phase modulation under certain conditions. These
divergences are \Red{limited} by intraband relaxation parameters, and lead to a large optical response from a high quality sample; we
\Red{find} they are very robust with respect to variations in the details
of the band structure. \Red{To clearly track all of these effects, we analyze
gapped graphene, describing the electrons as massive Dirac fermions;
under the relaxation time approximation, we derive analytic expressions
for the third order conductivities, and identify the divergences that
arise in describing the associated nonlinear phenomena.}
\end{abstract}
\maketitle

\section{Introduction}

Motivated by the novel optical properties of graphene\cite{Nat.Photon._4_611_2010_Bonaccorso,ACSNano_8_1086_2014_Low,Proc.R.Soc.A_473_20170433_2017_Ooi},
many researchers have turned their attention to the linear and nonlinear
optical response of 2D systems more generally \cite{Nat.Photon._10_227_2016_Sun,Adv.Mater._x_1705963_2018_Autere}.
While there are certainly strong-field excitation circumstances under
which a perturbative treatment will fail \cite{Science_356_736_2017_Yoshikawa,arXiv:1703.10945,Phys.Rev.B_95_035405_2017_Dimitrovski,Phys.Rev.B_96_195420_2017_Taucer},
for many materials a useful first step towards understanding the optical
response is the calculation of the conductivities that arise in an
expansion of the response of the induced current density in powers
of the electric field \cite{boyd_nonlinearoptics}. In materials where
inversion symmetry is present, or its lack can be neglected, the first
non-vanishing nonlinear response coefficient in the long-wavelength
limit arises at third order, and that is our focus in this paper.
The simplest approach one can take to calculate such response coefficients
is to treat the electrons in an independent particle approximation
\cite{Phys.Rev.B_52_14636_1995_Aversa}, describing any electron-electron
scattering effects and interactions with phonons by the introduction
of phenomenological relaxation rates. Such a strategy certainly has
its limitations, but at least it identifies many of the qualitative
features of the optical response, and in particular it identifies
what we call ``divergences'' in that response. \Red{We use this term
to refer to the infinite optical response coefficients that are
  predicted at certain frequencies or sets of frequencies in the
  so-called ``clean limit,'' where all scattering effects,
    including carrier-carrier scattering, 
    carrier-phonon scattering, and carrier-impurity scattering, are
    ignored by omitting any relaxation rates from the calculation}. Under these conditions
    the \Red{actual} predicted magnitude of a response depends critically on the values
chosen for the phenomenological relaxation rates. These ``divergences''
are of particular interest to experimentalists because they indicate
situations where the optical response can be expected to be large;
they are also of particular interest to theorists since they indicate
conditions under which a more sophisticated treatment of scattering
within the material, or perhaps a treatment of the response more sophisticated
than the perturbative one, is clearly in order.

The optical response of a crystal arises due to interband and intraband transitions \cite{Phys.Rev.B_52_14636_1995_Aversa}.
Resonances can be associated with both transitions. For linear optical response, only a single optical transition is
involved. A single interband transition can be on resonance for a
large range of photon energies, as long as the photon energy is above
the band gap. But the resonant electronic states are limited to those
with the energy difference matching the photon energy, which depend
on the details of the band structure. A single intraband transition
can be on resonance only for zero photon energy, and for electronic
states at the Fermi surface. Thus, whether or not these resonances
lead to a divergent optical response can \Red{depend strongly on the material being
considered.} For the nonlinear optical response, intraband and interband
transitions can be combined, leading to complicated nonlinear optical
transitions \cite{LightSci.App._5_16131_2016_Zhao,Phys.Rev.B_97_165111_2018_Soh}.
As with single intraband transitions, when the nonlinear optical
transitions involve the same initial and final electronic states the resonant
frequency, which is the sum of all involved frequencies, is also zero.
\Red{This is
analogous to the Drude conductivity in linear response, which diverges
at zero frequency in the ``clean limit.''} 
However, due to the interplay of interband transitions, the incident
frequencies need not necessarily all \Red{be} zero for there to be a divergence,
and the involved electronic states need not necessarily be around
the Fermi surface. \Red{By explicitly deriving the general expressions for
the third order nonlinear conductivities in the clean limit, we show
that the existence and characteristics of such divergences are of
a more general nature. To highlight them in a clear and tractable
way, we apply our approach to 2D gapped graphene, for which the perturbative
third order conductivities can be analytically obtained from the Dirac-like
band structure in the single particle approximation.} Although our
discussion is in the context of such 2D systems, the underlying physics
is the same for systems of different dimension.

Because the nonlinear transitions involve a number of frequencies,
these divergences can be classified into different types, associated
with different types of nonlinear phenomena. Several of these 
have been widely studied in \Red{the} literature, usually within the context
of a particular material or model or excitation condition; \Red{yet} the
connection with the general nonlinear conductivities is seldom discussed.
Our goal here is to demonstrate the \Red{general nature of the expressions
  for the response across a range of materials}.

The first type of divergence can be called ``current-induced second order nonlinearity''
(CISNL). It arises when free electrons in the system are driven by
an applied DC field; the induced DC current breaks the initial inversion
symmetry, and thus \Red{the material exihibits} an effective second-order response to
applied optical fields, leading to phenomena such as sum and difference
frequency generation. The nature of the divergence here is in the
response to the DC field, similar to a single intraband resonance,
which would be infinite if phenomenological relaxation terms were
not introduced; however, when written as proportional to the induced
DC current\Red{,} the effective second-order response coefficients are finite.
This phenomenon has been investigated extensively in different materials,
both experimentally \cite{Phys.Rev.B_85_121413_2012_Bykov,NanoLett._13_2104_2013_An,Phys.Rev.B_89_115310_2014_An,Phys.Rep._535_101_2014_Glazov,NanoLett._15_5653_2015_Yu}
and theoretically \cite{Opt.Express_22_15868_2014_Cheng,NanoLett._12_2032_2012_Wu,Appl.Phys.Lett._67_1113_1995_Khurgin}
. A second type is ``coherent current injection'' (CCI) \cite{CoherentControl_Driel,CoherentControlOfPhotocurrentsinSemiconductors_Driel,PhysicaE_45_1_2012_Rioux},
where the presence of fields at $\omega$ and $2\omega-\delta$ leads
to a divergent DC response as $\delta\rightarrow0$ if the excitation
at $2\omega$ is able to create free carriers; the divergent response
signals the injection of current by the interference of one-photon
absorption and degenerate two-photon absorption amplitudes. This is
the most widely studied process, both experimentally \cite{NanoLett._10_1293_2010_Sun,Phys.Rev.B_85_165427_2012_Sun}
and theoretically \cite{Phys.Rev.B_83_195406_2011_Rioux,Phys.Rev.B_93_075442_2016_Salazar,Phys.Rev.B_91_85404_2015_Muniz,Phys.Rev.B_86_115427_2012_Rao}.
Recent theoretical work has also identified an injection process associated with one-photon
absorption and the stimulated Raman process \cite{Phys.Rev.B_90_115424_2014_Rioux,Phys.Rev.B_93_075442_2016_Salazar}.
A third type is the jerk current
\cite{arXiv:1804.03970,arXiv_1806.01206}, \Red{which is a new type of one
  color CCI with the assistance of a static electric
  field. It} is a high order divergence involving both a static electric
field and an optical field. The static DC field can change the carrier
injection rate induced by the optical field, as well as a hydrodynamic
acceleration of these optically injected carriers; \Red{thus, as
  opposed to the usual two-color CCI, the injection rate of
the jerk current increases with the injection time.}

We can also identify new divergences, which have not been well recognized
in the literature, for two familiar third order nonlinear phenomena.
The first arises in cross-phase modulation (XPM) when fields at $\omega_{p}$
and $\omega_{s}$ are present. The response for the field at $\omega_{s}$
due to the field at $\omega_{p}$ can diverge when $\omega_{p}$ is
above or near the energy gap, leading to a phase modulation of the
field at $\omega_{s}$ that is limited by a relaxation rate. The second
also involves excitation with fields at $\omega_{p}$ and $\omega_{s}$,
but focuses on the degenerate four-wave mixing (DFWM) field generated
at $2\omega_{p}-\omega_{s}$. As $\omega_{s}\rightarrow\omega_{p}$
this term diverges for $\omega_{p}$ above or near the energy gap.
These cases merge as $\omega_{p}\to\omega_{s}$, which corresponds
to the most widely studied nonlinear phenomenon of Kerr effects and
two-photon absorption \cite{Opt.Express_18_4564_2010_Xing,NanoLett._11_5159_2011_Wu,NanoLett._11_2622_2011_Yang,Opt.Lett._37_1856_2012_Zhang,Nat.Photon._6_554_2012_Gu,Phys.Rev.Appl._6_044006_2016_Vermeulen,Opt.Lett._41_3281_2016_Dremetsika,Nat.Photon.___2018_Jiang,NC_Nathalie}.
The very large variation of the extracted values of the nonlinear
susceptibilities associated with these phenomena \cite{Nat.Photon.___2018_Jiang,Adv.Mater._x_1705963_2018_Autere,OpticalPropertiesofGraphene_RolfBinder}
may be related to such divergences.

In section II we review the general expressions for the third order
optical response in the independent particle approximation, and identify
in general the divergences that appear associated with the nonlinear
optical transitions with a vanishing total frequency. In section III
we specialize to the case of gapped graphene, and use it as an example
to illustrate the divergences. In section IV we point out the differences
between the divergent behavior of gapped and ungapped graphene. In
section V we conclude.

\section{The third order response \Red{conductivities}}

The general third order nonlinear susceptibility have been well studied
in \Red{the} literature for a cold intrinsic semiconductor \cite{Phys.Rev.B_52_14636_1995_Aversa},
with a large effort devoted to working out many subtle features. In
this section we mainly repeat the same procedure for a general band
system, and classify the expression in a way that the divergent term
can be easily identified.

Writing the electric field $\boldsymbol{E}(t)$ as 
\begin{align}
 & \boldsymbol{E}(t)=\sum_{i}\boldsymbol{E}(\omega_{i})e^{-i\omega t},
\end{align}
and other fields similarly, to third order in the electric field the
induced current density $\boldsymbol{J}^{(3)}(t)$ is characterized
by the response coefficients $\sigma^{(3);dabc}(\omega_{1},\omega_{2},\omega_{3})$,
\begin{align}
 & J^{(3);d}(\omega_{1}+\omega_{2}+\omega_{3})=\sum_{\omega_{1},\omega_{2},\omega_{3}}\sigma^{(3);dabc}(\omega_{1},\omega_{2},\omega_{3})E^{a}(\omega_{1})E^{b}(\omega_{2})E^{c}(\omega_{3})
\end{align}
where superscripts $a,b,...$ indicate Cartesian components and are
summed over when repeated. The coefficients $\sigma^{(3);abcd}(\omega_{1},\omega_{2},\omega_{3})$
can be taken to be symmetric under simultaneous permutation of $(bcd)$
and $(\omega_{1},\omega_{2},\omega_{3})$, and since the sums over
the $\omega_{i}$ are over all frequencies there are ``degeneracy
factors'' that arises under certain combinations of frequencies.
For example, if fields at $\omega_{p}$ and $\omega_{s}$ are present
we have 
\begin{align}
 & J^{d}(2\omega_{p}-\omega_{s})=3\sigma^{(3);dabc}(\omega_{p},\omega_{p},-\omega_{s})E^{a}(\omega_{p})E^{b}(\omega_{p})E^{c}(-\omega_{s}).
\end{align}
Often the response coefficient $\sigma^{(3);dabc}(\omega_{1},\omega_{2},\omega_{3})$
is written as $\sigma^{(3);dabc}(-\omega_{1}-\omega_{2}-\omega_{3};\omega_{1},\omega_{2},\omega_{3})$.
We do not do that here to avoid cluttering the notation, but we consider
it implicit in that when we picture these response coefficients we
draw arrows associated with all four of the variables appearing in
$\sigma^{(3);dabc}(-\omega_{1}-\omega_{2}-\omega_{3};\omega_{1},\omega_{2},\omega_{3})$,
upward arrows associated with positive variables and downward arrows
associated with negative variables.

To calculate the response coefficients in the independent particle
approximation we label the bands by lower case letters $n,m,$ etc.,
and the wave vectors in the first Brillouin zone by $\boldsymbol{k}$.
The density operator elements $\rho_{nm\boldsymbol{k}}$ associated
with bands $n,m$ and wave vector $\boldsymbol{k}$ satisfy the equation
of motion \cite{Phys.Rev.B_52_14636_1995_Aversa} 
\begin{align}
i\hbar\frac{\partial\rho_{nm\bm{k}}(t)}{\partial t} & =(\varepsilon_{n\bm{k}}-\varepsilon_{m\bm{k}})\rho_{nm\bm{k}}-e\bm{E}(t)\cdot\left[\sum_{l\neq n}\bm{\xi}_{nl\bm{k}}\rho_{lm\bm{k}}(t)-\sum_{l\neq m}\rho_{nl\bm{k}}(t)\bm{\xi}_{lm\bm{k}}\right]\nonumber \\
 & -ie\bm{E}(t)\cdot\left[\bm{\nabla}_{\bm{k}}\rho_{nm\bm{k}}(t)-i(\bm{\xi}_{nn\bm{k}}-\bm{\xi}_{mm\bm{k}})\rho_{nm\bm{k}}(t)\right]\nonumber \\
 & +i\hbar\left.\frac{\partial\rho_{nm\bm{k}}}{\partial t}\right|_{\text{scat}}\,.\label{eq:kbe0}
\end{align}
Here we describe the interaction of light with the matter using the
``$\boldsymbol{r\cdot}\boldsymbol{E}$'' approach rather than the
``$\boldsymbol{p\cdot}\boldsymbol{A}$'' approach involving the
vector potential $\boldsymbol{A}(t)$, for the latter can lead to
false divergences associated with the violation of sum rules when
the number of bands are inevitably truncated to make any calculation.
The coefficients $\boldsymbol{\xi}_{nm\boldsymbol{k}}$ are the Berry
connections, using the definition in the work by Aversa and Sipe \cite{Phys.Rev.B_52_14636_1995_Aversa}.
The interband optical transitions are identified by the off-diagonal
terms of $\bm{\xi}_{nm\bm{k}}$, while the rest of terms associated
with $\bm{E}(t)$ are associated with the intraband optical transitions.
The last term, \Red{$i\hbar\left.\frac{\partial\rho_{nm\bm{k}}}{\partial t}\right|_{\text{scat}}$},
describes the relaxation processes. In our approach, we take a relaxation
time approximation, and specify different relaxation time for different
transitions, as given below. We solve Eq.~(\ref{eq:kbe0}) perturbatively
by setting $\rho(t)=\sum\limits _{j\ge0}\rho^{(j)}(t)$, where $\rho^{(0)}(t)=\rho^{0}$
stands for the density matrix in the thermal equilibrium, and $\rho^{(j)}(t)\propto[\bm{E}]^{j}$.
The iteration for each order is given by 
\begin{align}
i\hbar\frac{\partial\rho_{nm\bm{k}}^{(j+1)}(t)}{\partial t} & =(\varepsilon_{n\bm{k}}-\varepsilon_{m\bm{k}})\rho_{nm\bm{k}}^{(j+1)}-e\bm{E}(t)\cdot\left[\sum_{l\neq n}\bm{\xi}_{nl\bm{k}}\rho_{lm\bm{k}}^{(j)}(t)-\sum_{l\neq m}\rho_{nl\bm{k}}^{(j)}(t)\bm{\xi}_{lm\bm{k}}\right]\nonumber \\
 & -ie\bm{E}(t)\cdot\left[\bm{\nabla}_{\bm{k}}\rho_{nm\bm{k}}^{(j)}(t)-i(\bm{\xi}_{nn\bm{k}}-\bm{\xi}_{mm\bm{k}})\rho_{nm\bm{k}}^{(j)}(t)\right]+i\hbar\left.\frac{\partial\rho_{nm\bm{k}}^{(j+1)}}{\partial t}\right|_{\text{scat}}\,,\label{eq:kbe1}
\end{align}
where we take the relaxation terms \cite{Phys.Rev.B_91_235320_2015_Cheng,*Phys.Rev.B_93_39904_2016_Cheng}
as 
\begin{align}
\hbar\left.\frac{\partial\rho_{nn\bm{k}}^{(j)}}{\partial t}\right|_{\text{scat}} & =-\Gamma_{a}^{(j)}\rho_{nn\bm{k}}^{(j)}\,,\\
\hbar\left.\frac{\partial\rho_{nm\bm{k}}^{(j)}}{\partial t}\right|_{\text{scat}} & =-\Gamma_{e}^{(j)}\rho_{nm\bm{k}}^{(j)}\,,\text{ for }n\neq m\,.
\end{align}
Here the $\Gamma_{e}^{(j)}$ and $\Gamma_{a}^{(j)}$ are phenomenological
relaxation parameters associated with interband and intraband motion
respectively, with the superscript $(j)$ indicating the order of
perturbation at which they are introduced. At optical frequencies,
the presence of relaxation parameters removes divergences associated
with the resonances, and the values of the relaxation parameters are
important for evaluating the nonlinear conductivities when resonant
transitions exist. It is natural to choose different values of $\Gamma_{e/a}^{(j)}$
for resonant and non-resonant transitions. A general perturbative
solution are presented in Appendix~\ref{app:pert}.

With the evolution of the density matrix, the current density can
be obtained through $\bm{J}(t)=e\sum_{nm}\int\frac{d\bm{k}}{(2\pi)^{2}}\bm{v}_{mn\bm{k}}\rho_{nm\bm{k}}(t)$,
where $\bm{v}_{mn\bm{k}}$ are the matrix elements for a velocity
operator. A third order response calculation \cite{boyd_nonlinearoptics}
leads to the result 
\begin{eqnarray}
\sigma^{(3);dabc}(\omega_{1},\omega_{2},\omega_{3}) & = & \frac{1}{6}\bigg[\widetilde{\sigma}^{(3);dabc}(\omega,\omega_{2}+\omega_{3},\omega_{3})+\widetilde{\sigma}^{(3);dacb}(\omega,\omega_{2}+\omega_{3},\omega_{2})\nonumber \\
 & + & \widetilde{\sigma}^{(3);dbac}(\omega,\omega_{1}+\omega_{3},\omega_{3})+\widetilde{\sigma}^{(3);dbca}(\omega,\omega_{1}+\omega_{3},\omega_{1})\nonumber \\
 & + & \widetilde{\sigma}^{(3);dcba}(\omega,\omega_{1}+\omega_{2},\omega_{1})+\widetilde{\sigma}^{(3);dcab}(\omega,\omega_{1}+\omega_{2},\omega_{2})\bigg]\,,
\end{eqnarray}
with $\omega=\omega_{1}+\omega_{2}+\omega_{3}$\Red{,} and the unsymmetrized
coefficients $\widetilde{\sigma}^{(3);dabc}(\omega,\omega_{0},\omega_{3})$
take the form 
\begin{eqnarray}
\widetilde{\sigma}^{(3);dabc}(\omega,\omega_{0},\omega_{3}) & = & \frac{1}{vv_{0}v_{3}}S_{1}^{dabc}+\frac{1}{vv_{0}}S_{2}^{dabc}(w_{3})+\frac{1}{vv_{3}}S_{3}^{dabc}(w_{0})+\frac{1}{v}S_{4}^{dabc}(w_{0},w_{3})\nonumber \\
 & + & \frac{1}{v_{0}v_{3}}S_{5}^{dabc}(w)+\frac{1}{v_{0}}S_{6}^{dabc}(w,w_{3})+\frac{1}{v_{3}}S_{7}^{dabc}(w,w_{0})+S_{8}^{dabc}(w,w_{0},w_{3})\,,\label{eq:sigma_tilde}
\end{eqnarray}
with 
\begin{align}
v & =\hbar\omega+i\Gamma_{a}^{(3)}\,, & w & =\hbar\omega+i\Gamma_{e}^{(3)}\,,\\
v_{0} & =\hbar\omega_{0}+i\Gamma_{a}^{(2)}\,, & w_{0} & =\hbar\omega_{0}+i\Gamma_{e}^{(2)}\,,\\
v_{3} & =\hbar\omega_{3}+i\Gamma_{a}^{(1)}\,, & w_{3} & =\hbar\omega_{3}+i\Gamma_{e}^{(1)}\,.
\end{align}
  \begin{table}[!htpb]
    \centering %
    \begin{tabular}[t]{|l|c|c|c|}
      \hline\hline
      &$v_3\sim 0$ & $v_0\sim 0$ & $v\sim 0$\\
      \hline
     Conductivity
    & $\sigma^{(3)}(\omega_1,\omega_2,\delta)$&
     $\sigma^{(3)}(\omega_1,\omega,-\omega+\delta)$&
      $\sigma^{(3)}(\omega_1,\omega_2,-\omega_1-\omega_2+\delta)$ \\
      \hline
      Divergence
      & $[\hbar\delta + i\Gamma_a^{(1)}]^{-1}$
      &$[\hbar\delta + i\Gamma_a^{(2)}]^{-1}$ &
      $[\hbar\delta + i\Gamma_a^{(3)}]^{-1}$ \\ \hline
      Nonlinear   & CINSL & XPM, DFWM & CCI\\
      phenomena & jerk current & jerk current & \\
\hline
\multirow{2}{*}{$\begin{array}{l}\text{Divergent}\\ \text{contributions}
\end{array}$} & &&\\
&         \mbox{\includegraphics[width=2.5cm]{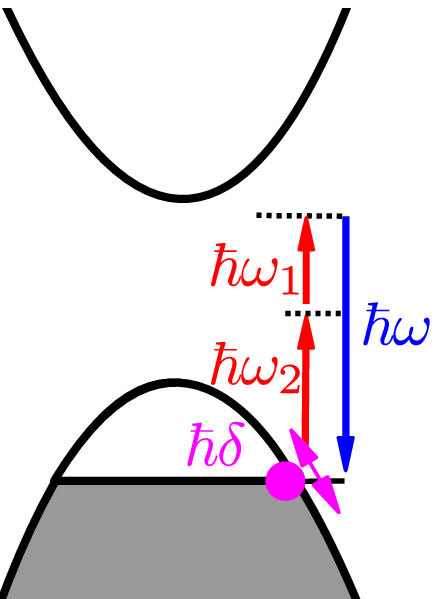}} &
         \mbox{\includegraphics[width=2.5cm]{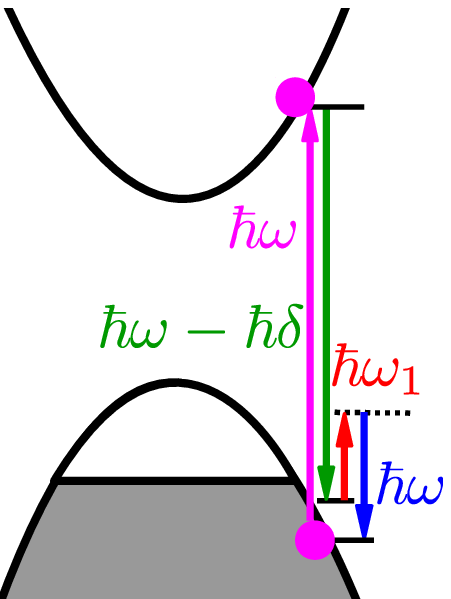}} &
         \mbox{\includegraphics[width=2.7cm]{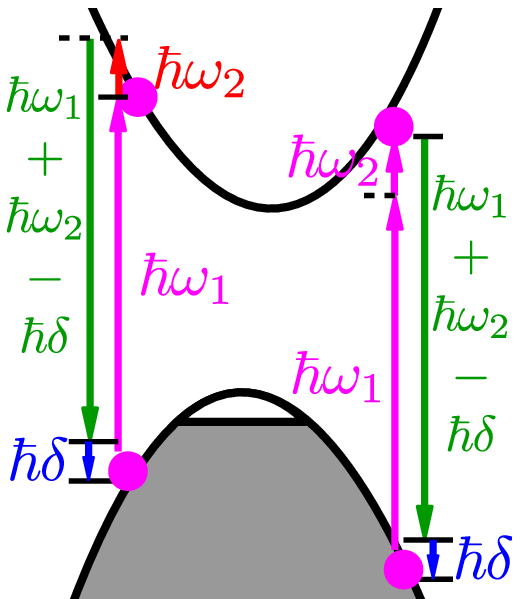}}\\
             \hline
$\begin{array}{l}\text{Resonant}\\ \text{conditions}
\end{array}$ & doped & $\hbar\omega>2E_c$ &
    $\begin{array}{l} (\text{left})~~~\hbar\omega_1>2E_c \\ (\text{right})~\hbar\omega_1+\hbar\omega_2>2E_c
    \end{array}$\\
    \hline\hline
    \end{tabular}
    \caption{Illustration of the resonant processes of of the
      divergences associated with the intraband
      motion. The row ``Divergent contributions''
      illustrates the divergent transitions, by the magenta arrows and
      labels in the diagram. The magenta dots show the doped or
      excited electronic states. The row ``Resonant conditions'' gives
      the  conditions for these resonances.} 
    \label{tab:sum1}    
  \end{table}
Note that the actual value of the energy appearing in, for example,
$v_{0}$ depends on the corresponding frequency (here $\omega_{0}$,
a sum of two of the incident frequencies) appearing in $\widetilde{\sigma}^{(3);dabc}(\omega,\omega_{0},\omega_{3})$.
The quantities $v$, $v_{0}$, and $v_{3}$ are associated with the
intraband motions (for carriers or excited carriers). The coefficients
$S_{i}^{dabc}$ are associated with interband transitions; we give
expressions for them, and for the expressions to which they reduce for the particular models we consider, in the Appendix~\ref{app:ghg}.
Any divergences they contain are associated with interband motion,
and thus all the intraband divergences are explicitly indicated by
the $v_{i}$ in the denominators appearing in Eq.~(\ref{eq:sigma_tilde}).
Thus it is the $\Gamma_{a}^{(i)}$ that will be of importance to us.
Typically one $\Gamma_{a}^{(i)}$ is important for a given divergence;
the other $\Gamma_{a}^{(j)}$, and the $\Gamma_{e}^{(j)}$ , to which
the process is not sensitive, are all set equal to a nominal value
$\Gamma$. \Red{These divergent processes are summarized in Table~\ref{tab:sum1}.}

\begin{table}[!htpb]
    \centering %
    \begin{tabular}[t]{|l|l|l|l|}
      \hline\hline
      Nonlinear &conductivity & Condition for & Note/Reference\\
      phenomenon & &   nonzero $A_i$ &\\
      \hline
      \multirow{2}{*}{CISNL} &
               \multirow{2}{*}{$\sigma^{(3)}(\omega_1,\omega_2,0)\rightarrow\frac{1}{\Gamma_a^{(1)}}A_1
                 + B_1$} & \multirow{2}{*}{doped} & $A_1$: CISHG {\cite{Appl.Phys.Lett._67_1113_1995_Khurgin,NanoLett._12_2032_2012_Wu,Opt.Express_22_15868_2014_Cheng,Phys.Rev.B_91_235320_2015_Cheng,*Phys.Rev.B_93_39904_2016_Cheng}{}}\\
               \cline{4-4}
               &&&$B_1$: EFISH {\cite{Phys.Rev.B_91_235320_2015_Cheng,*Phys.Rev.B_93_39904_2016_Cheng}{}}\\
      \hline
      XPM &
      $\sigma^{(3)}(\omega_s,\omega_p,-\omega_p)\rightarrow\frac{1}{\Gamma_a^{(2)}}
      A_1 + B_1 $ &  $\hbar\omega_p\ge 2E_c$ &\\
      \hline
      \multirow{2}{*}{DFWM$^{*}$} & \multirow{2}{*}{$\sigma^{(3)}(\omega_p,\omega_p,-\omega_s)\rightarrow\frac{1}{\Gamma_a^{(2)}}\left(\frac{1}{\Gamma} A_1 +
        A_2\right)+B_1$} & \multirow{2}{*}{$\hbar\omega_p\ge 2E_c$}  &
      $\omega_s\sim \omega_p$\\
        &&& $B_1$: Kerr
        effects \cite{Phys.Rev.B_91_235320_2015_Cheng,*Phys.Rev.B_93_39904_2016_Cheng}, TPA \\
        \hline
        \multirow{2}{*}{CCI} & \multirow{2}{*}{$\sigma^{(3)}(\omega,\omega,-2\omega)\rightarrow\frac{1}{\Gamma_a^{(3)}}(A_1 +A_2)+ B_1$} & $A_1:\quad \hbar\omega
        >~E_c$ & usual {CCI
          \cite{CoherentControl_Driel,CoherentControlOfPhotocurrentsinSemiconductors_Driel,PhysicaE_45_1_2012_Rioux,NanoLett._10_1293_2010_Sun,Phys.Rev.B_85_165427_2012_Sun,Phys.Rev.B_83_195406_2011_Rioux,Phys.Rev.B_93_075442_2016_Salazar,Phys.Rev.B_91_85404_2015_Muniz,Phys.Rev.B_86_115427_2012_Rao}{}}\\
        \cline{3-4}
        &  & $A_2:\quad \hbar\omega>2E_c$ &
        $\begin{array}{l}\text{stimulated Raman}\\\text{process \cite{Phys.Rev.B_90_115424_2014_Rioux,Phys.Rev.B_93_075442_2016_Salazar}{}} 
        \end{array}
        $\\
      \hline
      Jerk &
      \multirow{3}{*}{$\sigma^{(3)}(\omega,-\omega,0)\rightarrow \begin{array}{l}\frac{1}{\Gamma_a^{(3)}}\big(\frac{1}{\Gamma_a^{(2)}}A_1 +
      \frac{1}{\Gamma_a^{(1)}}A_2\\+A_3\big)+B_1 
        \end{array}
        $} & $A_1:\quad \hbar\omega>2E_c$&
      {\cite{arXiv:1804.03970,arXiv_1806.01206}{}}\\
      \cline{3-4}
      current&  &$A_2:\quad \text{doped}$&\\
      &&&\\
      \hline\hline
    \end{tabular}
    \caption{
      A short summary of the structure of the divergences in the nonlinear
optical phenomena discussed in this work. Here $2E_c$ is the
      gap or the chemical potential induced gap; all the $A_i$ and $B_i$
      are expansion coefficients. The third column
      lists the condition when the divergences can occur. (*) Here we set all the relaxation parameters $\Gamma_{a/e}^{(j)}$ except
      $\Gamma_{a}^{(2)}$ equal to $\Gamma$.}
    \label{tab:sum2}
  \end{table}
The general expression for the conductivity in
Eq.~(\ref{eq:sigma_tilde}) immediately indicates the possibilities of
the nonlinear phenomena discussed in the Introduction. For CISNL, the
conductivities 
$\sigma^{(3);dabc}(\omega_{1},\omega_{2},0)$ include divergences
associated with $v_{3}\to0$; for coherent current injection, the
conductivities $\sigma^{(3);dabc}(-2\omega,\omega,\omega)$ include
divergences associated with $v\to0$; the jerk current is a special
case of CISNL, described by $\sigma^{(3);dabc}(\omega,-\omega,0)$,
with divergences associated with $v_{3}\to0$ and $v\to0$; for XPM
and DFWM, the conductivities $\sigma^{(3);dabc}(\omega,\omega_{p},-\omega_{p})$
and $\sigma^{(3);dabc}(-\omega_{s},\omega_{p},\omega_{p})$ include
divergences associated with $v_{0}\to0$. In special cases, there
may be extra divergences identified by a combination of these limits,
and the detailed divergences types are determined by the values of
$S_{i}$. Of course, for finite relaxation times we will not have
a vanishing $v_{3}$, $v_{0}$, or $v$. Nonetheless, for frequencies
where the real part of one of these quantities vanishes the term(s)
in Eq.~(\ref{eq:sigma_tilde}) containing this quantity will make the
largest contribution, and we refer to them as the ``divergent contributions.''
Our focus is the identification of these divergent
contributions. \Red{Before getting into the details of these effects,
  it is helpful to isolate the divergent contributions in these
  conductivities, as shown in Table~\ref{tab:sum2}.} 

\section{Nonlinear optical conductivity \Red{of
gapped graphene}\label{sec:model}}

We apply the approach to gapped graphene, a two dimensional system.
The low energy excitations exist in two valleys, which can be described
by a simplified two band model for the unperturbed Hamiltonian 
\begin{eqnarray}
H_{\tau\bm{k}}^{0}(\Delta)=\begin{pmatrix}\Delta & \hbar v_{F}(ik_{x}+\tau k_{y})\\
\hbar v_{F}(-ik_{x}+\tau k_{y}) & -\Delta
\end{pmatrix}\,.\label{eq:h}
\end{eqnarray}
The quantity $\bm{k}=k_{x}\hat{\bm{x}}+k_{y}\hat{\bm{y}}$ is a two
dimensional wave vector, $\Delta\geq0$ is a mass parameter to induce
a band gap $2\Delta$, $\tau=\pm$ stands for a valley index, and
$v_{F}$ is the Fermi velocity. Ungapped graphene corresponds to the
limit $\Delta\rightarrow0$. All the necessary quantities for the
calculation of the third order conductivity are given \cite{Phys.Rev.B_92_235307_2015_Cheng} by
\begin{subequations}
\begin{align}
\varepsilon_{\tau s\bm{k}} & =s\epsilon_{k}\,,\text{ with }\epsilon_{k}=\sqrt{(\hbar v_{F}k)^{2}+\Delta^{2}}\,,\\
\bm{\xi}_{\tau+-\bm{k}} & =\frac{\hbar v_{F}(ik_{x}+\tau k_{y})}{2\epsilon_{\bm{k}}k^{2}}\left(-i\frac{\Delta}{\epsilon_{\bm{k}}}\bm{k}+\tau\bm{k}\times\hat{\bm{z}}\right)\,,\\
\bm{\xi}_{\tau++\bm{k}}-\bm{\xi}_{\tau--\bm{k}} & =\frac{1}{k^{2}}\left(1-\frac{\Delta}{\epsilon_{\bm{k}}}\right)\tau\bm{k}\times\hat{\bm{z}}\,,\\
\bm{v}_{\tau+-\bm{k}} & =2i\hbar^{-1}\epsilon_{\bm{k}}\xi_{\tau+-\bm{k}}\,,\\
\bm{v}_{\tau ss\bm{k}} & =\hbar^{-1}\bm{\nabla}_{\bm{k}}\varepsilon_{\tau s\bm{k}}\,.
\end{align}
\end{subequations} For the special two band system we use $s=\pm$
to indicate the upper ($+$) and lower ($-$) bands.

In this approximation for the dispersion relation of the bands, all
the $S_{j}^{dabc}$ can be analytically obtained, and are listed in
 Appendix~\ref{app:ghg}. Furthermore, the unsymmetrized conductivity
$\widetilde{\sigma}^{(3);dabc}(\omega,\omega_{0},\omega_{3})$ can
be written as the sum of two terms, 
\begin{align*}
 & \widetilde{\sigma}^{(3);dabc}(\omega,\omega_{0},\omega_{3})=\widetilde{\sigma}_{f}^{(3);dabc}(\omega,\omega_{0},\omega_{3})+\widetilde{\sigma}_{t}^{(3);dabc}(\omega,\omega_{0},\omega_{3}),
\end{align*}
where $\widetilde{\sigma}_{f}^{(3);dabc}$ includes all terms in Eq.~(\ref{eq:sigma_tilde})
in which $v_{3}$ appears, and $\widetilde{\sigma}_{t}^{(3);dabc}$
includes all remaining terms. Note that $\widetilde{\sigma}_{f}^{(3);dabc}$
is non-zero only for a doped system. Induced currents in this model
flow in the plane in which the gapped graphene is assumed to lie,
which we take to be the
$(xy)$ plane; $\sigma^{(3);dabc}$, $\widetilde{\sigma}^{(3);dabc}$,
and $S_{j}^{dabc}$ are fourth rank tensors, each with 16 components,
and the independent components can be taken to be those with the components
$xxyy$, $xyxy$, and $xyyx$. The other nonzero components can be
obtained through $S_{j}^{xxxx}=S_{j}^{xxyy}+S_{j}^{xyxy}+S_{j}^{xyyx}$,
and the symmetry $\{x\leftrightarrow y\}$. We list the independent
nonzero components of these tensors, taking $S_{j}^{dabc}(\cdots)$
as an example, as a column vector
$S_{j}=\begin{bmatrix}S_{j}^{xxyy}\\S_{j}^{xyxy}\\S_{j}^{xyyx}\end{bmatrix}$.

For gapped graphene, the expressions for $S_{i}$ show features similar
to the corresponding expressions for graphene \cite{Phys.Rev.B_91_235320_2015_Cheng,*Phys.Rev.B_93_39904_2016_Cheng}:
They are functions of the effective gap parameter $E_{c}=\text{max}\{\Delta,|\mu|\}$
and $\Delta$, in addition to the energies $w$, $w_{0}$, and $w_{3}$.
In all these expressions, $E_{c}$ appears only in functions of $E_{c}^{-5}$,
$E_{c}^{-3}$, $E_{c}^{-1}$, ${\cal I}(E_{c};w)$, ${\cal H}(E_{c};w)$,
${\cal H}(E_{c};w_{0})$, ${\cal G}(E_{c};w)$, ${\cal G}(E_{c};w_{0})$,
and ${\cal G}(E_{c};w_{3})$, where 
\begin{eqnarray}
{\cal G}(x;w) & = & \ln\left|\frac{w+2x}{w-2x}\right|+i\left(\pi+\arctan\frac{\text{Re}[w]-2x}{\text{Im}[w]}-\arctan\frac{\text{Re}[w]+2x}{\text{Im}[w]}\right)\,,\\
{\cal H}(x;w) & = & \frac{1}{2x+w}+\frac{1}{2x-w}\,,\\
{\cal I}(x;w) & = & \frac{1}{(2x+w)^{2}}-\frac{1}{(2x-w)^{2}}\,.
\end{eqnarray}
We can write each $S_{j}^{dabc}$ as a linear combination of these
functions, with their arguments themselves being functions of $w$,
$w_{0}$, $w_{3}$, and $\Delta$. All terms can be expanded in even
orders of $\Delta$, particularly as functions proportional to $\Delta^{0}$,
$\Delta^{2}$, and $\Delta^{4}$, as shown in Appendix~\ref{app:ghg}.

With all these analytic expressions in hand, any third order
nonlinear conductivity can be calculated and studied directly. In
the following, we consider the intraband divergences that are of interest
here, by giving the leading contributions to the conductivities.

\subsection{Current-Induced Second order Nonlinearity}

\begin{figure}[!htpb]
  \centering 
  \includegraphics[width=5cm]{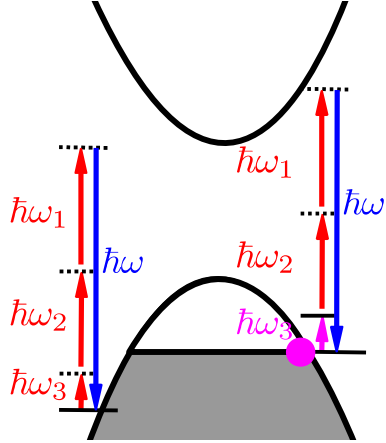}
  \caption{Illustration of the excitation scenario for CISNL. A general transition
process is sketched on the left hand side at a $\bm{k}$ point away
from the Fermi surface, and the intraband transition induced divergent
term (as $\hbar\omega_{3}\to0$) is sketched on the right hand side;
the latter occurs only for the electrons at the Fermi surface. The
short dashed horizontal lines stand for virtual states. }
\label{fig:cisl} 
\end{figure}

We begin by considering the response coefficient $\sigma^{(3)dabc}(\omega_{1},\omega_{2},0)$,
where we assume that neither $\omega_{1}$ nor $\omega_{2}$ vanishes,
and that their sum is finite. This describes a second-order optical
nonlinearity induced by a DC field ($\omega_{3}=0),$ and if there
are free carriers we would expect a divergent contribution because
of the large response of the free carriers to the DC field. A picture
of the excitation scenario is given in Fig.~\ref{fig:cisl}. In our
expression for $\widetilde{\sigma}^{(3);dabc}(\omega,\omega_{0},\omega_{3})$
the divergence of interest is clearly signaled by $v_{3}\rightarrow0$,
and the divergent response term can be obtained from the analytic
expressions. When this is isolated, the most important relaxation
parameter is $\Gamma_{a}^{(1)}$, which we treat differently than
we treat the other relaxation parameters; for $j=2,3$ we set \Red{$\Gamma_{a}^{(j)}=\Gamma_{e}^{(j)}=\Gamma$}
to give $v=w$ and $v_{0}=w_{0}$ in the expression for $\widetilde{\sigma}^{(3);dabc}$.
The divergent term can be written as 
\begin{equation}
\widetilde{\sigma}^{(3)}(\omega,\omega_{0},\omega_{3}\to0)\approx\frac{i\sigma_{3}}{\Delta^{5}}\frac{E_{c}^{2}-\Delta^{2}}{\hbar\omega_{3}+i\Gamma_{a}^{(1)}}{\cal Z}_{1}\left(\frac{w}{\Delta},\frac{w_{0}}{\Delta};\frac{E_{c}}{\Delta}\right)\,,\label{eq:cis-1}
\end{equation}
with $\sigma_{3}=\sigma_{0}(\hbar v_{F}e)^{2}/\pi$, $\sigma_{0}=e^{2}/(4\hbar)$,
and 
\begin{eqnarray}
{\cal Z}_{1}(x,x_{0};\alpha) & = & \frac{-16}{\alpha xx_{0}\left(x^{2}-4\alpha^{2}\right)^{2}\left(x_{0}^{2}-4\alpha^{2}\right)}\left\{ \begin{bmatrix}4\alpha^{2}x\left(2x+x_{0}\right)\\
-x^{3}x_{0}\\
x_{0}\left(x^{3}+x_{0}\left(3x^{2}-4\alpha^{2}\right)\right)
\end{bmatrix}\right.\nonumber \\
 & + & \left.\left[4\alpha^{2}\left(\alpha^{2}+3\right)-\left(3\alpha^{2}+1\right)x^{2}-\left(x^{2}+4\right)x_{0}^{2}-2\left(\alpha^{2}+1\right)x_{0}x\right]\begin{bmatrix}1\\
1\\
1
\end{bmatrix}\right\} \,.\label{eq:Z1_expression}
\end{eqnarray}
These expressions have a number of interesting features: (1) The divergent
term in Eq.~(\ref{eq:cis-1}) does not include any contributions from the
${\cal G}$ functions, indicating immediately that it is the existence
of free carriers that is important. Indeed, this can be immediately
confirmed, for if the system is undoped we have $E_{c}=\Delta$, and
the term in Eq.~(\ref{eq:cis-1}) vanishes. (2) The divergent term is inversely
proportional to the relaxation parameter. Since in a very rough approximation
the DC current satisfies \Red{$\bm{J}_{\text{DC}}\propto\left(\Gamma_a^{(1)}\right)^{-1}\boldsymbol{E}_{\text{DC}}$}
with the DC field $\boldsymbol{E}_{\text{DC}}$, the divergent contribution
to $\sigma^{(3)dabc}(\omega_{1},\omega_{2},0)$ can be written as
a finite term proportional to $\boldsymbol{J}_{\text{DC}}$, hence
the identification of this response as ``current-induced second order
nonlinearity.'' (3) The expression in Eq.~(\ref{eq:Z1_expression}) for
${\cal Z}_{1}(x,x_{0};\alpha)$ can exhibit further divergences as
$w\to2E_{c}$ and $w_{0}\to2E_{c}$, indicating that interband divergences
can arise in CISNL as well, depending on the values of the optical
frequencies $\omega_{1}$ and $\omega_{2}$. 
\begin{figure}[!htpb]
\centering 
\includegraphics[width=15cm]{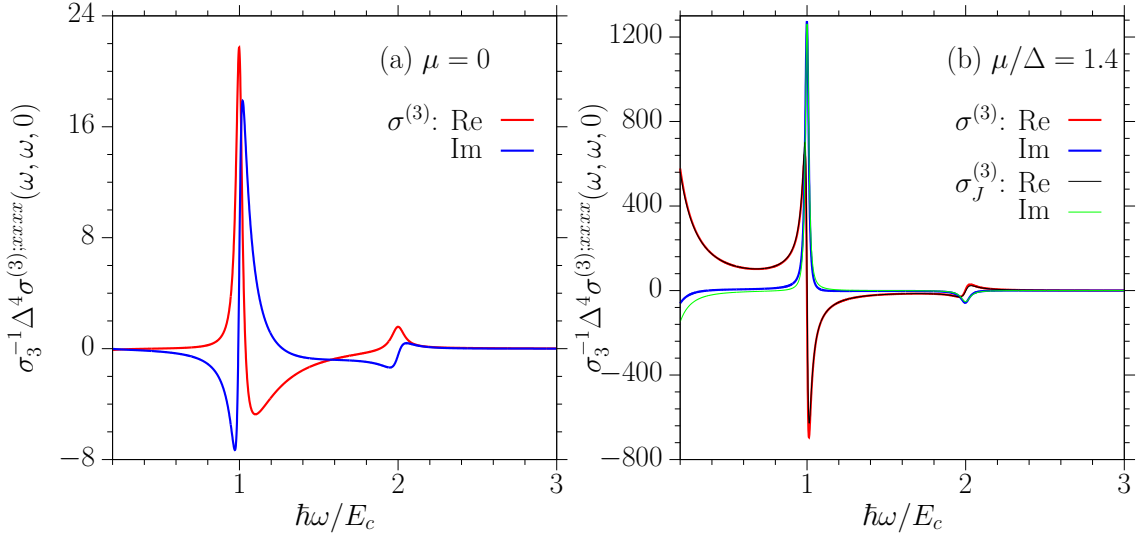}
\caption{Conductivity for second harmonic generation (a) $\mu=0$ and (b) $\mu/\Delta=1.4$
for $\Gamma/\Delta=0.05$ and $\Gamma_{a}^{(3)}/\Delta=0.01$. In
(b), the current induced contribution is also plotted.}
\label{fig:cshg} 
\end{figure}

We refer to the terms in $\sigma^{(3)dabc}(\omega_{1},\omega_{2},0)$
that are $not$ divergent as $v_{3}\rightarrow0$ as the ``field-induced
second order nonlinearity.'' They exist in the absence of free carriers,
and have an analogue in the field induced second order nonlinearity
of usual semiconductors, which leads to processes such as electric-field
induced second harmonic generation (EFISH) \cite{Phys.Rev.B_91_235320_2015_Cheng,*Phys.Rev.B_93_39904_2016_Cheng,SolidStateCommun._246_76_2016_Margulis}.
Adopting this perspective, we can write the effective second order
conductivity $\sigma^{(2);dab}(\omega_{1},\omega_{2})=3\sigma^{(3);dabc}(\omega_{1},\omega_{2},0)E_{\text{DC}}^{c}$
with 
\begin{equation}
\sigma^{(3);dabc}(\omega_{1},\omega_{2},0)=\sigma_{J}^{(3);dabc}(\omega_{1},\omega_{2},0)+\sigma_{E}^{(3);dabc}(\omega_{1},\omega_{2},0)\,.
\end{equation}
The first term characterizes the current-induced second order response
coefficient, and arises from the divergent contribution in Eq.~(\ref{eq:cis-1}),
\begin{eqnarray}
\sigma_{J}^{(3);dabc}(\omega_{1},\omega_{2},0) & = & \frac{\sigma_{3}E_{\text{dc}}^{c}}{6\Gamma_{a}^{(1)}}\left[\left(\frac{E_{c}}{\Delta}\right)^{2}-1\right]\left[{\cal Z}_{1}^{dabc}\left(\frac{\hbar(\omega_{1}+\omega_{2})+i\Gamma}{\Delta},\frac{\hbar\omega_{2}+i\Gamma}{\Delta};\frac{E_{c}}{\Delta}\right)\notag\right.\\
 &  & \left.+{\cal Z}_{1}^{dbac}\left(\frac{\hbar(\omega_{1}+\omega_{2})+i\Gamma}{\Delta},\frac{\hbar\omega_{1}+i\Gamma}{\Delta};\frac{E_{c}}{\Delta}\right)\right]\,.
\end{eqnarray}
All the remaining contributions are collected in the field induced
response coefficient $\sigma_{E}^{(3);dabc}(\omega_{1},\omega_{2},0)$.

Figure~\ref{fig:cshg} shows the spectrum of $\sigma^{(3);xxxx}(\omega,\omega,0)$
with a relaxation parameter $\Gamma/\Delta=0.05$ for (a) an undoped
system with no free carriers, $\mu/\Delta=0$, and (b) a doped system
with $\mu/\Delta=1.4$. For both systems, there are obvious resonant
peaks at $\hbar\omega=nE_{c}$ for $n=1,2$ associated with interband
transitions; in the response of the doped system, there is an additional
peak as $\hbar\omega\rightarrow0$ , and here the contribution from
the current-induced SHG dominates for photon energies away from the
resonances. This divergent process results in a qualitatively larger
conductivity $\sigma^{(3)dabc}(\omega,\omega,0)$ for a doped system
(b) than for an undoped system (a).

\subsection{Cross-Phase Modulation}

\begin{figure}[!htpb]
  \centering 
  \includegraphics[width=6cm]{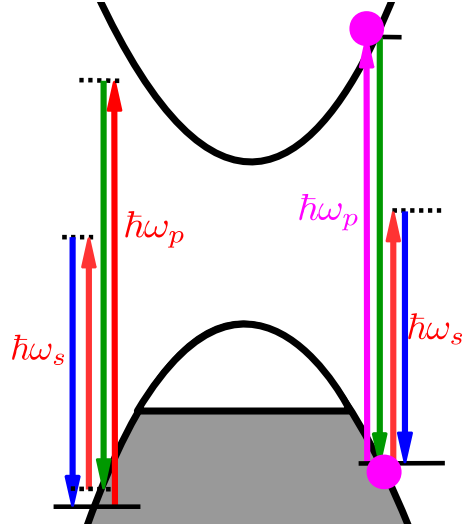}
  \caption{Illustration of the excitation scenario for XPM. Similar to Fig.~\ref{fig:cisl},
the transition at the left hand side is off resonance, while the one
at the right is on resonance.}
\label{fig:xpm} 
\end{figure}

In cross-phase modulation the propagation of light at frequency $\omega_{s}$
is modified by the presence of an intense optical field at a different
frequency, $\omega_{p}$; it is characterized by the response coefficient
$\sigma^{(3)}(\omega_{s},-\omega_{p},\omega_{p})$. This conductivity
exhibits a divergence as $w_{0}\to0$. Unlike CISNL, the divergence
arises from the nonlinear response of the system and a static field
is not required. The process is pictured in Fig.~\ref{fig:xpm};
we will see below that the divergent term vanishes unless $\hbar\omega_{p}$
is near or above the effective band gap $2E_{c}$. Cross-phase modulation
of a signal frequency $\omega_{s}$ can be effected not just by a
CW beam at $\omega_{p}$ , but by a pulse of light centered at such
a frequency. So in general we seek $\sigma^{(3)}(\omega_{s},-\omega_{p}+\delta,\omega_{p})$,
where knowledge of this expression for a small range of frequencies
$\omega_{p}$ centered about a nominal pump frequency, and for a small
range of detunings $\delta$ centered around zero, will allow us calculated
the cross-phase modulation of a signal by a pump pulse. As $\delta,\Gamma_{a}^{(2)}\to0$,
the leading term of the relevant unsymmetrized conductivity is 
\begin{equation}
\widetilde{\sigma}^{(3)}(\omega_{s}+\delta,\delta\to0,\omega_{p})\approx\frac{i\sigma_{3}}{\Delta^{3}}\frac{2}{\hbar\delta+i\Gamma_{a}^{(2)}}{\cal Z}_{2}\left(\frac{w}{\Delta},\frac{w_{3}}{\Delta};\frac{E_{c}}{\Delta}\right)\,,\label{eq:limitv0}
\end{equation}
with \Red{$w=v=\hbar(\omega_{s}+\delta)+i\Gamma$, $w_{3}=v_{3}=\hbar\omega_{p}+i\Gamma$},
where we set all first order and third order relaxation energies to
be the same for simplicity, and 
\begin{eqnarray}
{\cal Z}_{2}(x,x_{3};\alpha) & = & \frac{8(A_{0}+2xx_{3}A_{4})}{x^{3}x_{3}^{3}\alpha}+\frac{8A_{0}}{3x^{3}x_{3}\alpha^{3}}+\frac{(x^{2}-4)^{2}A_{0}+4x^{2}(4-x^{2})A_{5}}{2x^{4}x_{3}}{\cal I}(\alpha;x)\nonumber \\
 & + & \frac{(x^{2}-4)^{2}A_{0}+32x^{2}A_{5}}{2x^{5}x_{3}}{\cal H}(\alpha;x)\nonumber \\
 & + & Z_{1}(x,x_{3}){\cal G}(\alpha;x)+Z_{2}(x,x_{3}){\cal G}(\alpha;x_{3})\,.\label{eq:cpmcalz-1}
\end{eqnarray}
and 
\begin{eqnarray}
Z_{1}(x,x_{3}) & = & \frac{x_{3}(x^{2}-4)^{2}A_{0}+32x^{2}(x_{3}A_{5}+xA_{4})}{2x^{6}(x^{2}-x_{3}^{2})}\,,\label{eq:cpmz1-1}\\
Z_{2}(x,x_{3}) & = & -\frac{x(x_{3}^{2}-4)^{2}A_{0}+32xx_{3}(x_{3}A_{5}+xA_{4})}{2x^{2}x_{3}^{4}(x^{2}-x_{3}^{2})}\,.\label{eq:cpmz2-1}
\end{eqnarray}
Here we used $A_0=-(A_1+A_2+A_3)$, $A_{4}=\frac{1}{4}(A_{2}-A_{3})$ and $A_{5}=-\frac{1}{4}(2A_{1}+A_{2}+A_{3})$,
where 
\begin{align}
A_{1} & =\begin{bmatrix}-3\\
1\\
1
\end{bmatrix}\,,\quad A_{2}=\begin{bmatrix}1\\
-3\\
1
\end{bmatrix}\,,\quad A_{3}=\begin{bmatrix}1\\
1\\
-3
\end{bmatrix}\,.\label{eq:defA}
\end{align}
We note that $Z_{1}$ and $Z_{2}$ themselves diverge
as $x\to-x_{3}$ (i.e., $\omega_{p}\rightarrow\omega_{s})$, which
will be discussed in next section.

Using the expression above we can write 
\begin{equation}
\sigma^{(3)}(\omega_{s},-\omega_{p}+\delta,\omega_{p})=\sigma_{\text{xpm};d}^{(3)}(\omega_{s},\omega_{p};\delta)+\sigma_{\text{xpm}}^{(3)}(\omega_{s},\omega_{p};\delta)\,,
\end{equation}
where $\sigma_{\text{xpm};d}^{(3)}(\omega_{s},\omega_{p};\delta)$
contains a divergence with respect to $\left(\hbar\delta+i\Gamma_{a}^{(2)}\right)^{-1}$
\begin{eqnarray}
\sigma_{\text{xpm};d}^{(3);dabc}(\omega_{s},\omega_{p};\delta) & = & \frac{i\sigma_{3}}{3\Delta^{4}}\left(\frac{\hbar\delta+i\Gamma_{a}^{(2)}}{\Delta}\right)^{-1}\left[{\cal Z}_{2}^{(3);dabc}\left(\frac{\hbar\omega_{s}+i\Gamma}{\Delta},\frac{-\hbar\omega_{p}+i\Gamma}{\Delta};\frac{E_{c}}{\Delta}\right)\notag\right.\\
 &  & \left.+{\cal Z}_{2}^{(3);dacb}\left(\frac{\hbar\omega_{s}+i\Gamma}{\Delta},\frac{\hbar\omega_{p}+i\Gamma}{\Delta};\frac{E_{c}}{\Delta}\right)\right],
\end{eqnarray}
and $\sigma_{\text{xpm}}^{(3)}(\omega_{s},\omega_{p};\delta)$ contains
the non-divergent response. Note that the divergent term is independent
of the sequences of the limits $\delta\to0$ and $\Gamma\to0$.

\begin{figure}[!htpb]
  \centering
  \includegraphics[width=14.4cm]{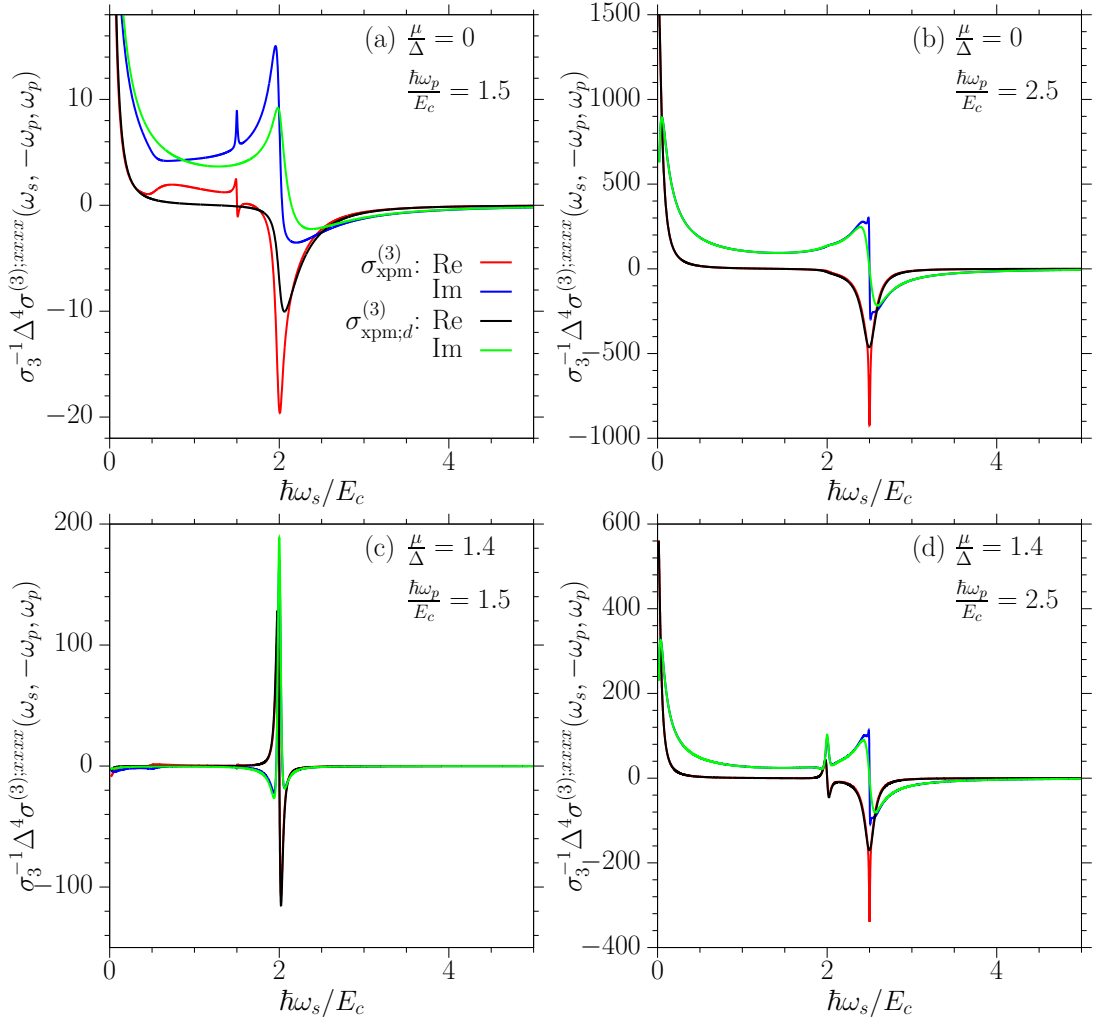}
\caption{Signal frequency $\omega_{s}$ dependence of the conductivity $\sigma_{3}^{-1}\Delta^{4}\sigma_{\text{xpm}}^{(3);xxxx}(\omega_{s},\omega_{p};0)$
at different chemical potential and pump frequencies for $\Gamma/\Delta=0.05$
and $\Gamma_{a}^{(2)}/\Delta=0.01$. The parameters of $\left(\frac{\mu}{\Delta},\frac{\hbar\omega_{p}}{E_{c}}\right)$
are (a) $(0,1.5)$, (b) $(0,2.5)$, (c) $(1.4,1.5)$, and (d) $(1.4,2.5)$.
The divergent contribution $\sigma_{\text{xpm};d}^{(3);xxxx}(\omega_{s},\omega_{p};0)$
are also shown for comparison.}
\label{fig:xpm3} 
\end{figure}

Figure~\ref{fig:xpm3} gives the spectrum of $\sigma_{\text{xpm}}^{(3);xxxx}(\omega_{s},\omega_{p};0)$,
appropriate for CW illumination with the pump; as above we have taken
a relaxation parameter $\Gamma/\Delta=0.05$ and $\Gamma_{a}^{(2)}/\Delta=0.01$,
and again consider an undoped system $\left(\mu/\Delta=0\right)$
and a doped system with $\mu/\Delta=1.4$. The complicated dependence
of both the real and imaginary parts on the frequencies indicates
the rich nature of the nonlinear response. The real parts show additional
divergences as $\hbar\omega_{s}\to0$ and $\hbar\omega_{p}\to0$,
as would be expected from the discussion of CISNL above, and as $\hbar\omega_{s}\to\hbar\omega_{p}$,
as we discuss in the section below. As would be expected from the
discussion of CISNL above, there are other divergences or resonant
peaks associated with either $\hbar\omega_{s}\to0$ or interband transitions;
the latter are not our focus in this work. Away from these resonances,
the values of the conductivity for the cases $\hbar\omega_{p}/E_{c}=1.5$
are much smaller than those for the cases $\hbar\omega_{p}/E_{c}=2.5$,
which include the intraband divergences, and the contribution from
$\sigma_{\text{xpm};d}^{(3);xxxx}(\omega_{s},\omega_{p};0)$ is generally
the largest part of $\sigma^{(3)}(\omega_{s},-\omega_{p},\omega_{p})$
away from these other divergences. We can get some insight into the
importance of pump frequency by noting that as $\Gamma\rightarrow0$
the divergent term $\sigma_{\text{xpm};d}^{(3);dabc}(\omega_{s},\omega_{p};\delta)$
becomes 
\begin{equation}
\sigma_{\text{xpm};d}^{(3)}(\omega_{s},\omega_{p};\delta)\rightarrow\frac{i\sigma_{3}}{\Delta^{4}}\left(\frac{\hbar\delta+i\Gamma_{a}^{(2)}}{\Delta}\right)^{-1}h_{1}\left(\frac{\hbar\omega_{p}}{\Delta},\frac{\hbar\omega_{s}}{\Delta}\right)T\left(\frac{E_{c}}{\Delta},\frac{\hbar\omega_{p}}{\Delta}\right)\label{eq:cleanxpm}
\end{equation}
with 
\begin{equation}
  h_{1}(x,x_{1})=\frac{(x^{2}-4)^{2}A_{0}+32x(xA_{5}+x_{1}A_{4})}{6x^{4}x_{1}(x^{2}-x_{1}^{2})}
\label{eq:xpmh}
\end{equation}
and 
\begin{align}
 & T(y,x)=\mathcal{G}(y;x+i0^{+})+\mathcal{G}(y;-x+i0^{+})=2i\pi\theta(x-2y).\label{eq:clean_T}
\end{align}
So in the clean limit\Red{, where all $\Gamma_{a/e}^{(j)}\to 0$}, $\sigma_{\text{xpm};d}^{(3)}(\omega_{s},\omega_{p};\delta)$
vanishes if $\hbar\omega_{p}$ is below the effective gap $2E_{c}$. 
The frequency variation $\delta$ comes from the pumping beam,
which is negligible for very long pulse duration; then
$\sigma_{\text{xpm};d}^{(3)}$ 
is \Red{pure imaginary, and} the modulation is \Red{associated with } refraction rather
than absorption. Deviations from this arise from the finite
relaxation rate for other transition processes, but clearly
this divergence is associated with resonant excitation from the valence
to conduction band at $\hbar\omega_{p}$, as sketched in Fig.~\ref{fig:xpm}.

\subsection{Degenerate Four Wave Mixing}

\begin{figure}[!htpb]
  \centering 
  \includegraphics[width=4cm]{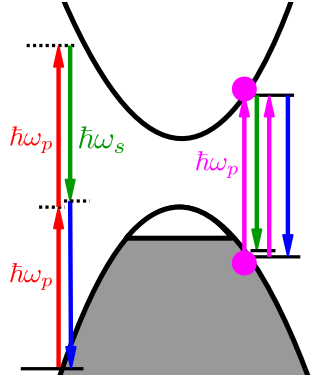}
  \caption{Illustration of the excitation scenario for degeneration FWM. Similar
to Fig.~\ref{fig:cisl}, the transition at the left hand side is
off resonance, while the one at the right is on resonance.}
\label{fig:dfwm} 
\end{figure}

In Eqs.~(\ref{eq:cpmcalz-1}) and (\ref{eq:xpmh}) there appear to
be divergences as $v\to\pm v_{3}$. Working out the expression in
detail it can be immediately seen that in fact no divergence results
as $v\to v_{3}$, but a divergence does indeed arise as $v\to-v_{3}$.
The leading divergence is associated with terms of the form $v_{0}^{-1}(v+v_{3})^{-1}$.
It gives a higher order divergence with a Lorentz type lineshape,
as shown in Figs.~\ref{fig:xpm3} (b) and (d). Such \Red{a} divergence also
exists in \Red{the} widely studied nonlinear phenomena of four-wave mixing,
which is characterized by the response coefficient $\sigma^{(3)}(\omega_{p},\omega_{p},-\omega_{s})$.
As for XPM, the divergent point is at $\omega_{s}=\omega_{p}$, but
since the generated field is at $2\omega_{p}-\omega_{s}$
rather than at $\omega_{s}$ the path to the divergence is different.

As $\omega_{s}\to\omega_{p}$, the divergent term arises in the unsymmetrized
conductivities $\widetilde{\sigma}^{(3)}(2\omega_{p}-\omega_{s},\omega_{p}-\omega_{s},-\omega_{s})$
and $\widetilde{\sigma}^{(3)}(2\omega_{p}-\omega_{s},\omega_{p}-\omega_{s},\omega_{p})$.
Taking $\delta=\omega_{s}-\omega_{p}$ and all $\Gamma_{a/e}^{(j)}$
as small quantities, the conductivity can be approximated as 
\begin{eqnarray}
{\sigma}^{(3)}(\omega_{p},\omega_{p},-\omega_{s})&\approx&\frac{i\sigma_{3}}{\Delta^{4}}\frac{\Delta}{\hbar\delta+i\Gamma_{a}^{(2)}}\left[\frac{\Delta}{\hbar\delta+i\Gamma}{\cal
    Z}_{3}\left(\frac{\hbar\omega_{p}}{\Delta},\frac{\Gamma}{\Delta};\frac{E_c}{\Delta}\right)\right.\notag\\
  && \left.+{\cal
    Z}_{4}\left(\frac{\hbar\omega_{p}}{\Delta},\frac{\Gamma}{\Delta};\frac{E_c}{\Delta}\right)\right]\,.
\end{eqnarray}
Here we set all the relaxation parameters $\Gamma_{a/e}^{(j)}$ except
$\Gamma_{a}^{(2)}$ equal to $\Gamma$. The exact expression can be
obtained following Eq.~(\ref{eq:limitv0}), or even from the full
expression in Appendix.~\ref{app:ghg}. The physics of this divergence
can be revealed if we consider the clean limit for ${\cal Z}_{3}$
and ${\cal Z}_{4}$. They are ${\cal
  Z}_j(x,0;\alpha)=Z_j(x)T(\alpha,x)$ for $j=3, 4$ with
\begin{align}
Z_{3}(x) & =\frac{1}{12x^{6}}\left[(x^{4}+24x^{2}+16)A_{0}-48x^{2}A_{6}\right]\,\\
Z_{4}(x) & =\frac{1}{12x^{7}}\left[(x^{4}-56x^{2}-48)A_{0}+160x^{2}A_{6}\right]\,,
\end{align}
with $A_{6}=-(A_{1}+A_{2}+2A_{3})/4$. 
Here $T(\alpha;x)$ survives only for $x>2\alpha$ (see Eq.~(\ref{eq:clean_T})),
where one photon absorption exists. Thus a relevant picture for this
kind of resonance is shown in Fig.~\ref{fig:dfwm}.

\begin{figure}[!htpb]
\centering 
\includegraphics[width=15cm]{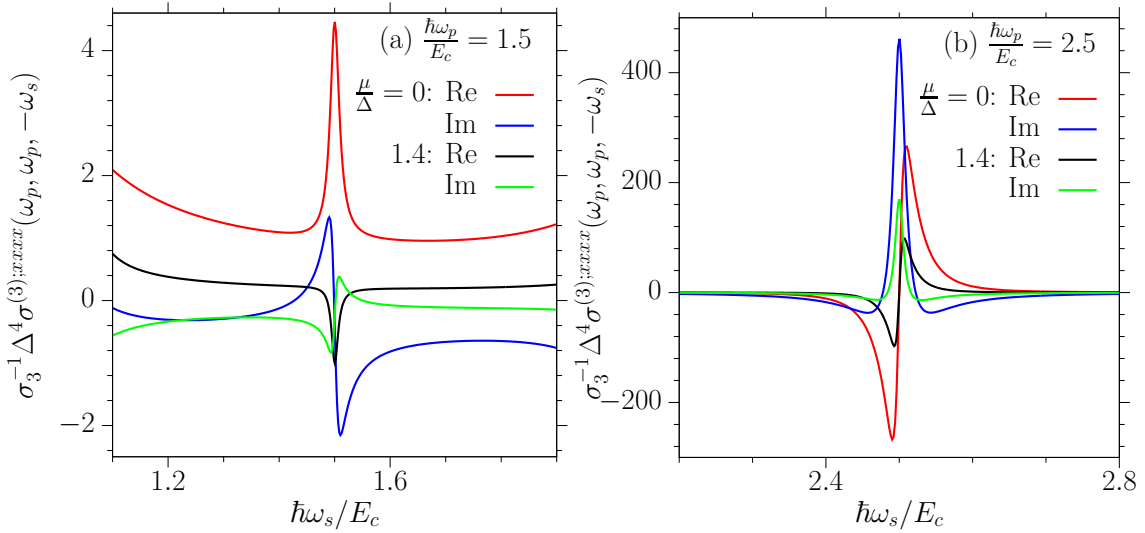}
\caption{Conductivity ${\sigma}^{(3);xxxx}(\omega_{p},\omega_{p},-\omega_{s})$
for $\omega_{s}$ around $\omega_{p}$ at different chemical potential
$|\mu|/\Delta=0$ and $1.4$ and different pump frequency $\hbar\omega_{p}/E_{c}=1.5$
and $2.5$. The relaxation parameters are taken as $\Gamma_{a}^{(2)}/\Delta=0.01$
and $\Gamma/\Delta=0.05$, \Red{and} the gap parameter is taken as $\Delta=1$~eV. }
\label{fig:fwm} 
\end{figure}

We illustrate these divergences in Fig.~\ref{fig:fwm} for two different
pump frequencies, $\hbar\omega_{p}/E_{c}=1.5$ and $2.5$. We plot
the results for an undoped system, $|\mu|/\Delta=0$ , and a doped
system with $\mu|/\Delta=1.4$. The conductivity strongly depends
on the ratio ${\hbar\omega}/{E_{c}}$. For $\hbar\omega/(2E_{c})<1$,
the divergent term vanishes in the clean limit and only the non-divergent
term survives. The features in Fig.~\ref{fig:fwm}(a) as $\hbar\omega_{s}$
is around $\hbar\omega_{p}$ are induced by the nonzero relaxation
parameters. For $\hbar\omega/(2E_{c})>1$ the divergent term survives
and to leading order varies as $\left[\hbar\delta+i\Gamma_{a}^{(2)}\right]^{-1}(\hbar\delta+i\Gamma)^{-1}$,
where we separate out the second order intraband relaxation parameter,
and put the rest equal to $\Gamma$. The near degenerate FWM is approximately
determined by 
\begin{eqnarray*}
\sigma^{(3)}(\omega_{p},\omega_{p},-\omega_{p}+\delta) &
\approx&-\frac{2\pi\sigma_{3}}{\Delta^{2}[\hbar\delta+i\Gamma_{a}^{(2)}](\hbar\delta+i\Gamma)}Z_{3}\left(\frac{\hbar\omega_p}{\Delta}\right)\notag\\
&&\times\theta\left(\hbar\omega_p-2E_{c}\right)\,.
\end{eqnarray*}
Note that the nonlinear response to a pulse of light can expected
to be very complicated due to the strong dependence on the detuning.

\subsection{Coherent Current Injection}

\begin{figure}[!htpb]
  \centering 
  \includegraphics[width=4cm]{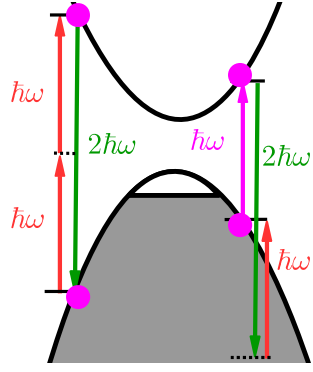}
  \caption{Illustration of the excitation scenario for XPM. The transition at
the left hand side corresponds to the interference between 
one-photon absorption and degenerate two-photon
absorption, while the transition at the
right hand side corresponds to the interference between one-photon
absorption and the stimulated Raman process.}
\label{fig:cci} 
\end{figure}

Now we turn to another divergence associated with $v\to0$ in Eq.~(\ref{eq:sigma_tilde}),
which exists in the conductivity $\sigma^{(3)}(-2\omega+\delta,\omega,\omega)$
as $\delta\to0$. This divergence describes coherent current injection.
Setting all the relaxation parameters $\Gamma_{a/e}^{(j)}$ except
$\Gamma_{a}^{(3)}$ equal to $\Gamma$, we have the leading contribution
with respect to the small quantity $\Gamma_{a}^{(3)}$ and $\delta$
given by 
\begin{equation}
\sigma^{(3)}(-2\omega+\delta,\omega,\omega)=\frac{\eta_{\text{cci}}(\omega,\Gamma)}{\hbar^{-1}\Gamma_{a}^{(3)}-i\delta}+\cdots\label{eq:sigmacci}
\end{equation}
While the full expressions of $\eta_{\text{cci}}(\omega,\Gamma)$
can be obtained from our general analytic expressions, the underlying
physics can be easily shown at the clean limit; setting $\Gamma\to0^{+}$,
we find 
\begin{eqnarray}
\eta_{\text{cci}}(\omega,0) & = & \frac{\sigma_{3}}{\hbar\Delta^{3}}\left[g_{1}\left(\frac{\hbar\omega}{\Delta}\right)T\left(\frac{E_{c}}{\Delta},\frac{2\hbar\omega}{\Delta}\right)+g_{2}\left(\frac{\hbar\omega}{\Delta}\right)T\left(\frac{E_{c}}{\Delta},\frac{\hbar\omega}{\Delta}\right)\right]\,,
\end{eqnarray}
where the functions $g_{1}$ and $g_{2}$ are given by 
\begin{eqnarray}
g_{1}(x) & = & \frac{1}{3x^{7}}\begin{bmatrix}(1-x^{2})^{2}\,, & 1-x^{4}\,, & 1-x^{4}\end{bmatrix}^{T}\,,\\
g_{2}(x) & = & \frac{1}{12x^{7}}\begin{bmatrix}-16+24x^{2}-5x^{4}\,, & -16-8x^{2}+3x^{4}\,, & -16-8x^{2}+3x^{4}\end{bmatrix}^{T}\,.\label{eq:cci_g}
\end{eqnarray}
The coefficient $\eta_{\text{cci}}$ describes the two-color coherent
current injection coefficient. In the clean limit, it includes two
terms: the one involving $g_{2}$ is nonzero only for $\hbar\omega>2E_{c}$,
and the one involving $g_{1}$ is nonzero for $\hbar\omega>E_{c}$.
Both terms arises because of the interference of one-photon and two-photon
absorption processes. However, the contribution from the term involving
$g_{1}$ comes from the interference between the pathways of one-photon
absorption and degenerate two-photon
absorption, as illustrated in the transition
process on the left side of Fig.~\ref{fig:cci}\Red{. While} the term involving
$g_{2}$ comes from the interference between one-photon absorption
and the stimulated Raman process, as shown in the transition process
on the right side of Fig.~\ref{fig:cci}. In experiments involving
typical semiconductors $2\hbar\omega$ is usually greater than the
band gap energy but $\hbar\omega$ is not; however, if $\hbar\omega$
also is greater than the gap there can be a contribution due to stimulated
electronic Raman scattering. In the clean limit, a direct calculation
of the functions $g_{i}(x)$ gives $g_{1}(1)=g_{2}(2)=0$. For the
component $\sigma^{(3);xxxx}$, their contributions are maximized
at $x\approx1.2$ for $g_{1}(x)$ and $x\approx2.4$ for $g_{2}(x)$.
However, at $x\approx2.4$, the contribution from the stimulated electronic
Raman scattering is no more than 20\% of the contribution from the
usual injection process; the total contribution has no maximum around
$x\approx2.4$.

A direct consequence of the divergence $v\to0$ is the form of the current response to a pulse light. For simplicity, we only take
the beam at $2\omega$ as a pulse, and then the time evolution of
the current density is 
\begin{align}
J^{(3);d}(t)=\int\frac{d\delta}{2\pi}\sigma^{(3);dabc}(-2\omega+\delta,\omega,\omega)E^{a}(-2\omega+\delta)E^{b}(\omega)E^{c}(\omega)e^{-i\delta t}\,.
\end{align}
With substitution of the leading term in Eq.~(\ref{eq:sigmacci}) we can get 
\begin{equation}
\frac{dJ^{(3);d}(t)}{dt}=\eta_{cci}^{dabc}(\omega,\Gamma)E_{2\omega}^{a}(t)E_{\omega}^{b}(t)E_{\omega}^{c}(t)-\hbar^{-1}\Gamma_{a}^{(3)}J^{(3);d}(t)\,.
\end{equation}
Here $\bm{E}_{2\omega}(t)$ and $\bm{E}_{\omega}(t)$ are the time
evolution of pulses with center frequencies at $2\omega$ and $\omega$,
respectively. The first term at the right hand side is a source term,
while the second term describes the damping. This equation shows exactly
how the injection process occurs.

\begin{figure}[!htpb]
\centering 
 \includegraphics[width=15cm]{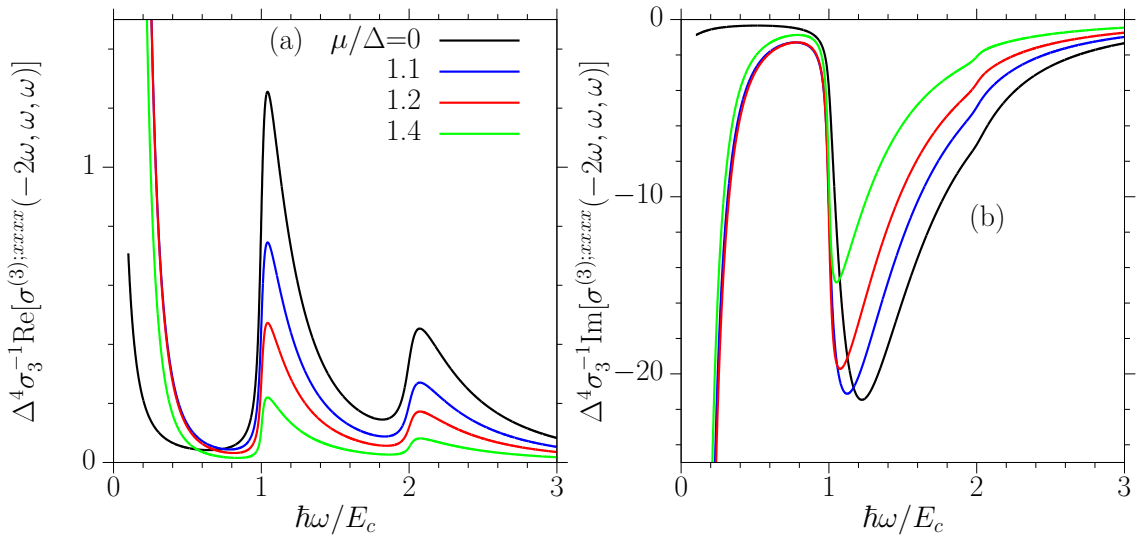}
\caption{Conductivity ${\sigma}^{(3);xxxx}(-2\omega,\omega,\omega)$ for different
chemical potential $|\mu|/\Delta=0$, $1.1$, $1.2$, and $1.4$.
The relaxation parameters are taken as $\Gamma_{a}^{(3)}/\Delta=0.01$
and $\Gamma/\Delta=0.05$, the gap parameter is taken as $\Delta=1$~eV. }
\label{fig:ci} 
\end{figure}

As an illustration, we plot the ${\sigma}^{(3);xxxx}(-2\omega,\omega,\omega)$
as a function of $\omega$ for different chemical potentials $|\mu|/\Delta=0$,
$1.1$, $1.2$, and $1.4$; the other parameters are taken as $\Gamma_{a}^{(3)}/\Delta=0.01$,
$\Gamma/\Delta=0.05$, and $\Delta=1$~eV. In the clean limit, ${\sigma}^{(3);xxxx}$
is \Red{purely} imaginary, and although with the inclusion of damping the
real parts do not vanish, they are about an order of magnitude smaller
than the imaginary parts, as shown in Fig.~\ref{fig:ci}. In the
calculations at finite 
relaxation parameters, both the real and the imaginary parts show
obvious peaks/valleys around $\hbar\omega\sim E_{c}$ and $\hbar\omega\sim2E_{c}$,
which correspond to the contributions from the terms including $g_{1}$
and $g_{2}$, respectively. In contrast to the situation discussed after Eq.~(\ref{eq:cci_g}) for clean limits, the appearance of
the peaks for $\hbar\omega>2E_{c}$
arises because of the
inclusion of the finite relaxation parameters. There also exists
increases of the conductivity values as $\hbar\omega\to0$, mostly
for nonzero $\mu/\Delta$. They are associated with a divergence induced
by the free-carriers, which is not our focus in this work.

\subsection{Jerk Current}

Some of the divergences considered here can appear simultaneously.
A good example is the recently discussed jerk current
\cite{arXiv:1804.03970}, which can be treated as a special case of XPM for a zero signal frequency,
or a special case of CISNL for current induced one-photon current
injection. The corresponding conductivity is $\sigma^{(3);dabc}(\omega,-\omega,0)$,
which involves the unsymmetrized conductivities $\widetilde{\sigma}^{(3);dabc}(0,-\omega,0)$,
$\widetilde{\sigma}^{(3);dacb}(0,-\omega,-\omega)$, $\widetilde{\sigma}^{(3);dbac}(0,\omega,0)$,
$\widetilde{\sigma}^{(3);dbca}(0,\omega,\omega)$, $\widetilde{\sigma}^{(3);dcab}(0,0,-\omega)$,
and $\widetilde{\sigma}^{(3);dcba}(0,\omega,0)$. They all include
intraband divergences, and the highest order is described by the limiting
behavior as $v,v_{0}\to0$
or $v,v_{3}\to0$. In general, the leading orders are 
\begin{align}
\sigma^{(3);dabc}(\omega,-\omega,0) & \approx\frac{i\sigma_{3}}{\Delta^{4}}\frac{\Delta}{i\Gamma_{a}^{(3)}}\left[\frac{\Delta}{i\Gamma_{a}^{(2)}}Q_{1}\left(\frac{\hbar\omega}{\Delta};\left\{ \frac{\Gamma_{a/e}^{(j)}}{\Delta}\right\} \right)+\frac{\Delta}{i\Gamma_{a}^{(1)}}Q_{2}\left(\frac{\hbar\omega}{\Delta};\left\{ \frac{\Gamma_{a/e}^{(j)}}{\Delta}\right\} \right)\right]\,.
\end{align}
The clean limits of $Q_{1}$ and $Q_{2}$ are 
\begin{align}
Q_{1}(x;0) & =\frac{1}{6x^{6}}\left[(x^{2}-4)^{2}A_{0}+32x^{2}A_{6}\right]T(\alpha;x)\,,\\
Q_{2}(x;0) & =\frac{(\alpha^{2}-1)(\alpha^{2}+3)}{3\alpha^{5}}\frac{1}{x}A_{0}\,,
\end{align}
with $\alpha=E_{c}/\Delta$. Here $Q_{1}(x;0)$ is nonzero only when
the one-photon absorption exists as $\hbar\omega>2E_{c}$, while $Q_{2}(x;0)$
exists only for a doped system where $\alpha>1$. These two terms have a different power dependence, and the term
involving $Q_{2}(x;0)$ can exists for any optical field frequency.
Thus the frequency dependence of the response can be used to distinguish
between them. 

\section{Comparison with graphene}

Some of the phenomena discussed above have been considered earlier
for doped graphene \cite{Opt.Express_22_15868_2014_Cheng,Phys.Rev.B_91_235320_2015_Cheng,*Phys.Rev.B_93_39904_2016_Cheng,NewJ.Phys._16_53014_2014_Cheng,*Corrigendum_NewJ.Phys._18_29501_2016_Cheng}.
Although the expressions we derived above are normalized to the gap
parameter $\Delta$, it is safe to take the limit as $\Delta\to0$.
This is because our general expressions of the conductivity can be
safely reduced to the case of graphene with taking $\Delta=0$, as
shown in Appendix~\ref{app:ghg}. The results of CISNL, DFWM, and
CCI in graphene have been discussed earlier \cite{Opt.Express_22_15868_2014_Cheng,Phys.Rev.B_91_235320_2015_Cheng,*Phys.Rev.B_93_39904_2016_Cheng,NewJ.Phys._16_53014_2014_Cheng,*Corrigendum_NewJ.Phys._18_29501_2016_Cheng}
in the clean limit and with finite relaxation parameters. Here we
give a brief discussion for XPM and the jerk current in graphene.
In the clean limit, the XPM for graphene can be found from Eq.~(\ref{eq:cleanxpm})
to be 
\begin{align}
\sigma_{\text{xpm};d}^{(3)}(\omega_{s},\omega_{p};\delta)\rightarrow\frac{i\sigma_{3}}{\hbar\delta}\frac{T\left(|\mu|;\hbar\omega_{p}\right)}{12\hbar\omega_{s}[(\hbar\omega_{p})^{2}-(\hbar\omega_{s})^{2}]}A_{0}\,,\label{eq:cleanxpmgh}
\end{align}
Here the chemical potential induced gap plays a role similar to that
of the gap parameter in gapped graphene. For the jerk current, the
leading term becomes 
\begin{align}
\sigma^{(3);dabc}(\omega,-\omega,0) & \to\frac{\sigma_{3}}{3i\Gamma_{a}^{(3)}}\left[\frac{1}{\Gamma_{a}^{(2)}}\frac{T(|\mu|;\hbar\omega)}{2(\hbar\omega)^{2}}+\frac{1}{\Gamma_{a}^{(1)}}\frac{1}{|\mu|(\hbar\omega)}\right]A_{0}\,.
\end{align}

\section{Conclusions}

We have systematically discussed intraband divergences in the
third order optical response, and identified the leading terms in
the corresponding third order conductivity. Due \Red{to} the combination
of intraband and interband transitions, these divergences can appear
at optical frequencies, and lead to large nonlinear conductivities. We have shown that the existence of such
divergences is very general, independent of the details of the band
structure. We illustrated these divergences in gapped graphene, \Red{with 
analytic expressions obtained for the third order conductivities in the  
relaxation time approximation.}

Such divergences are of interest to experimentalists, because within
the independent particle treatment presented here the optical response
is limited only by the phenomenological relaxation times introduced
in the theory, and thus that optical response can be expected to be
large. As well, at a qualitative level the predicted nature of the
divergent behavior is robust against approximations made in describing
the details of the interband transitions. The divergences are also
of interest to theorists, because one can expect that under such conditions
the kind of treatment presented here is too naive. This could be both
because more realistic treatments of relaxation processes are required,
and as well because the large optical response predicted could be
an indication that in a real experimental scenario the perturbative
approach itself is insufficient. Thus the identification of these
divergences identifies regions of parameter space where experimental
and theoretical studies can be expected to lead to new insights into
the nature of the interaction of light with matter.

More generally, we can expect that the calculation of other
response coefficients involving perturbative expressions of the density
matrix response to an electric field will reveal similar divergences
in the nonlinear contributions
to the response of other physical quantities, such as carrier density, spin/valley polarization, and spin/valley current. A deep understanding of these divergences can lead to new ways to
probe these quantities, and to study new effects in the optical response
of materials that depend on them. 

\acknowledgments This work has been supported by CAS QYZDB-SSW-SYS038,
NSFC Grant No. 11774340 and 61705227. S.W.W. is supported by the
National Basic Research Program of China under Grant 
No. 2014CB921601S. J.E.S. is supported by the Natural Sciences and
Engineering Research Council of Canada.

\appendix

\section{Perturbative conductivity for general band structure\label{app:perp}\label{app:pert}}

In this appendix we give the formal derivation of the third order
conductivities in terms of the electron energy, velocity matrix elements,
and Berry connections. The density matrix can be expanded in terms
of electric fields as 
\begin{align}
\rho_{nm\bm{k}}(t) & =\sum_{\omega_{3}}\widetilde{{\cal P}}_{nm\bm{k}}^{(1);c}(\omega_{3})E^{c}(\omega_{3})e^{-i\omega_{3}t}\nonumber \\
 & +\sum_{\omega_{2}\omega_{3}}\widetilde{{\cal P}}_{nm\bm{k}}^{(2);bc}(\omega_{0},\omega_{3})E^{b}(\omega_{2})E^{c}(\omega_{3})e^{-i\omega_{0}t}\nonumber \\
 & +\sum_{\omega_{1}\omega_{2}\omega_{3}}\widetilde{{\cal P}}_{nm\bm{k}}^{(3);abc}(\omega,\omega_{0},\omega_{3})E^{a}(\omega_{1})E^{b}(\omega_{2})E^{c}(\omega_{3})e^{-i\omega t}+\cdots\,,
\end{align}
with $\omega=\omega_{1}+\omega_{2}+\omega_{3}$ and $\omega_{0}=\omega_{2}+\omega_{3}$.
Using the iteration in Eq.~(\ref{eq:kbe1}), we can get the density
matrix at different perturbation orders. The first order terms of
the density matrix are 
\begin{align}
\widetilde{{\cal P}}_{nn\bm{k}}^{(1);c}(\omega_{3}) & =\frac{1}{v_{3}}A_{1;nn\bm{k}}^{(1);c}\,, & A_{1;nn\bm{k}}^{(1);c} & =i\frac{\partial f_{n\bm{k}}}{\partial k_{a}}\,,\\
\widetilde{{\cal P}}_{nm\bm{k}}^{(1);c}(\omega_{3}) & =B_{1;nm\bm{k}}^{(1);c}(w_{3})\,, & B_{1;nm\bm{k}}^{(1);c}(w_{3}) & =\frac{r_{nm\bm{k}}^{a}f_{mn\bm{k}}}{w_{3}-(\varepsilon_{n\bm{k}}-\varepsilon_{m\bm{k}})}\,.
\end{align}
The second order terms are 
\begin{eqnarray}
\widetilde{{\cal P}}_{nn\bm{k}}^{(2);bc}(\omega_{2},\omega_{3}) & = & \frac{1}{v_{0}v_{3}}A_{1;nn\bm{k}}^{(2);bc}+\frac{1}{v_{0}}A_{2;nn\bm{k}}^{(3);bc}(w_{3})\,,\\
\widetilde{{\cal P}}_{nm\bm{k}}^{(2);bc}(\omega_{2},\omega_{3}) & = & \frac{1}{v_{3}}B_{1;nm\bm{k}}^{(2);bc}(w_{0})+B_{2;nm\bm{k}}^{(2);bc}(w_{0},w_{3})\,,
\end{eqnarray}
with 
\begin{eqnarray}
A_{1;nn\bm{k}}^{(2);bc} & = & i\frac{\partial A_{1;nn\bm{k}}^{(1);c}}{\partial k_{b}}\,,\\
A_{2;nn\bm{k}}^{(2);bc}(w_{3}) & = & \sum_{m}[r_{nm\bm{k}}^{b}B_{1;mn\bm{k}}^{(1);c}(w_{3})-B_{1;nm\bm{k}}^{(1);c}(w_{3})r_{mn\bm{k}}^{b}]\,,\\
B_{1;nm\bm{k}}^{(2);bc}(w_{0}) & = & \frac{r_{nm\bm{k}}^{b}[A_{1;mm\bm{k}}^{(1);c}-A_{1;nn\bm{k}}^{(1);c}]}{w_{0}-(\varepsilon_{n\bm{k}}-\varepsilon_{m\bm{k}})}\,,\\
B_{2;nm\bm{k}}^{(2);bc}(w_{0},w_{3}) & = & \frac{[r_{\bm{k}}^{b},B_{1;\bm{k}}^{(1);c}(w_{3})]_{nm}+i\left(B_{1;\bm{k}}^{(1);c}(w_{3})\right)_{;nmk_{b}}}{w_{0}-(\varepsilon_{n\bm{k}}-\varepsilon_{m\bm{k}})}\,.\label{eq:gi}
\end{eqnarray}
Here we have used the notation $\bm{r}_{nm\bm{k}}=(1-\delta_{nm})\bm{\xi}_{nm\bm{k}}$
and 
\begin{eqnarray}
\left(B_{1;\bm{k}}^{(1);c}(w_{3})\right)_{;nmk_{b}} & = & \frac{\partial B_{1;nm\bm{k}}^{(1);c}(w_{3})}{\partial k_{b}}-i(\xi_{nn\bm{k}}^{b}-\xi_{mm\bm{k}}^{b})B_{1;nm\bm{k}}^{(1);c}(w_{3})\,.
\end{eqnarray}
It shows that the diagonal terms of the Berry connections $\bm{\xi}_{nn\bm{k}}$
appear always with the derivative $\bm{\nabla}_{\bm{k}}$ to form
a gauge invariant term \cite{Phys.Rev.B_52_14636_1995_Aversa}. The
third order terms are 
\begin{eqnarray}
\widetilde{{\cal P}}_{nn\bm{k}}^{(3);abc}(\omega_{1},\omega_{2},\omega_{3}) & = & \frac{1}{vv_{0}v_{3}}A_{1;nn\bm{k}}^{(3);abc}+\frac{1}{vv_{0}}A_{2;nn\bm{k}}^{(3);abc}(w_{3})\nonumber \\
 & + & \frac{1}{vv_{3}}A_{3;nn\bm{k}}^{(3);abc}(w_{0})+\frac{1}{v}A_{4;nn\bm{k}}^{(3);abc}(w_{0},w_{3})\,,\\
\widetilde{{\cal P}}_{nm\bm{k}}^{(3);abc}(\omega_{1},\omega_{2},\omega_{3}) & = & \frac{1}{v_{0}v_{3}}B_{1;nm\bm{k}}^{(3);abc}(w)+\frac{1}{v_{0}}B_{2;nm\bm{k}}^{(3);abc}(w,w_{3})\nonumber \\
 & + & \frac{1}{v_{3}}B_{3;nm\bm{k}}^{(3);abc}(w,w_{0})+B_{4;nm\bm{k}}^{(3);abc}(w,w_{0},w_{3})\,,
\end{eqnarray}
with 
\begin{eqnarray}
A_{1;nn\bm{k}}^{(3);abc} & = & i\frac{\partial A_{1;nn\bm{k}}^{(2);bc}}{\partial k_{a}}\,,\\
A_{2;nn\bm{k}}^{(3);abc}(w_{3}) & = & i\frac{\partial A_{2;nn\bm{k}}^{(2);bc}(w_{3})}{\partial k_{a}}\,,\\
A_{3;nn\bm{k}}^{(3);abc}(w_{0}) & = & \sum_{m}[r_{nm\bm{k}}^{a}B_{1;mn\bm{k}}^{(2);bc}(w_{0})-B_{1;nm\bm{k}}^{(2);bc}(w_{0})r_{mn\bm{k}}^{a}]\,,\\
A_{4;nn\bm{k}}^{(3);abc}(w_{0},w_{3}) & = & \sum_{m}[r_{nm\bm{k}}^{a}B_{2;mn\bm{k}}^{(2);bc}(w_{0},w_{3})-B_{2;nm\bm{k}}^{(2);bc}(w_{0},w_{3})r_{mn\bm{k}}^{a}]\,,
\end{eqnarray}
and 
\begin{eqnarray}
B_{1;nm\bm{k}}^{(3);abc}(w) & = & \frac{r_{nm\bm{k}}^{b}[A_{1;mm\bm{k}}^{(2);bc}-A_{1;nn\bm{k}}^{(2);bc}]}{w-(\varepsilon_{n\bm{k}}-\varepsilon_{m\bm{k}})}\,,\\
B_{2;nm\bm{k}}^{(3);abc}(w,w_{3}) & = & \frac{r_{nm\bm{k}}^{b}[A_{2;mm\bm{k}}^{(2);bc}(w_{3})-A_{2;nn\bm{k}}^{(2);bc}(w_{3})]}{w-(\varepsilon_{n\bm{k}}-\varepsilon_{m\bm{k}})}\,,\\
B_{3;nm\bm{k}}^{(3);abc}(w,w_{0}) & = & \frac{[r_{\bm{k}}^{a},B_{1;\bm{k}}^{(2);bc}(w_{0})]_{nm}+i\left(B_{1;\bm{k}}^{(2);bc}(w_{0})\right)_{;nmk_{a}}
}{w-(\varepsilon_{n\bm{k}}-\varepsilon_{m\bm{k}})}\,,\\
B_{4;nm\bm{k}}^{(3);abc}(w,w_{0},w_{3}) & = & \frac{[r_{\bm{k}}^{a},B_{2;\bm{k}}^{(2);bc}(w_{0},w_{3})]_{nm}+i\left(B_{2;\bm{k}}^{(2);bc}(w_{0})\right)_{;nmk_{a}}
}{w-(\varepsilon_{n\bm{k}}-\varepsilon_{m\bm{k}})}
\end{eqnarray}

The current density is then calculated through $\bm{J}(t)=e\sum_{nm}\int\frac{d\bm{k}}{(2\pi)^{2}}v_{mn\bm{k}}^{d}\rho_{nm\bm{k}}(t)$,
and the $S_{j}$ terms are given by 
\begin{align}
S_{i}^{dabc}(\cdots) & =-e^{4}\sum_{n}\int\frac{d\bm{k}}{(2\pi)^{2}}v_{nn\bm{k}}^{a}A_{i;nn\bm{k}}^{(3);abc}(\cdots)\,,\text{ for }i=1,2,3,4,\\
S_{i}^{dabc}(\cdots) & =-e^{4}\sum_{nm}\int\frac{d\bm{k}}{(2\pi)^{2}}v_{mn\bm{k}}^{a}B_{i;nm\bm{k}}^{(3);abc}(\cdots)\,,\text{ for }i=5,6,7,8\,.
\end{align}

\section{$S_{j}^{dabc}$ for a gapped graphene\label{app:ghg}}

The calculation of $S_{j}^{dabc}$ is straightforward. In the linear
dispersion approximation and relaxation time approximation, analytic
expressions can be obtained. By listing the nonzero components of
$S_{j}^{dabc}(\cdots)$ as a column vector \Red{$S_{j}=\begin{bmatrix}S_{j}^{xxyy} \\ S_{j}^{xyxy} \\ S_{j}^{xyyx}\end{bmatrix}$},
we can write 
\begin{align}
S_{j} & =i\sigma_{3}\left[W_{j}^{(0)}+2\Delta^{2}W_{j}^{(2)}+2\Delta^{4}W_{j}^{(4)}\right]\theta(|\mu|-\Delta)\,, & \text{ for }j=1,3,5,7\,,\nonumber \\
S_{j} & =i\sigma_{3}\left[W_{j}^{(0)}+2\Delta^{2}W_{j}^{(2)}+2\Delta^{4}W_{j}^{(4)}\right]\,, & \text{ for }j=2,4,6,8\,.
\end{align}
with\Red{
\begin{align}
  \sigma_{3}=\sigma_{0}\frac{(\hbar v_{F}e)^{2}}{\pi}\,, \quad \sigma_{0}=\frac{e^{2}}{4\hbar}\,.
\end{align}}
Note that $W^{(j)}(\cdots)$ depends on these functions 
\begin{eqnarray*}
\begin{array}{rrr}
|E_{c}|^{-5}\,, & |E_{c}|^{-3}\,, & |E_{c}|^{-1}\,,\notag\\
{\cal I}(E_{c};w)\,, & {\cal H}(E_{c};w)\,, & {\cal H}(E_{c};w_{0})\,,\\
{\cal G}(E_{c};w)\,, & {\cal G}(E_{c};w_{0})\,, & {\cal G}(E_{c};w_{3})\,.
\end{array}
\end{eqnarray*}
These functions depend on the photon energies and the effective gap
parameter $E_{c}$.

\subsection{$W_{j}^{(0)}$}
\Red{Gapped graphene reduces} to graphene \cite{Phys.Rev.B_91_235320_2015_Cheng,*Phys.Rev.B_93_39904_2016_Cheng}
as $\Delta\to0$ and $E_{c}\to|\mu|$, and thus $W_{j}^{(0)}$ should
give the results for graphene, as \begin{subequations} 
\begin{eqnarray}
W_{1}^{(0)} & = & \frac{1}{E_{c}}A_{0}\,,\label{eq:s1}\\
W_{3}^{(0)}(w_{0}) & = & {\cal H}(E_{c};w_{0})\frac{A_{3}}{w_{0}}-\frac{1}{E_{c}}\frac{A_{3}}{w_{0}}\,,\\
W_{5}^{(0)}(w) & = & {\cal H}(E_{c};w)\frac{A_{0}}{w}+{\cal I}(E_{c};w)A_{1}-\frac{1}{E_{c}}\frac{A_{0}}{w}\,,\\
W_{7}^{(0)}(w,w_{0}) & = & {\cal H}(E_{c};w_{0})\left(\frac{A_{2}}{w_{1}^{2}}-\frac{A_{3}}{w_{0}w_{1}}\right)+{\cal H}(E_{c};w)\left(\frac{A_{3}}{ww_{1}}-\frac{A_{2}}{w_{1}^{2}}\right)\nonumber \\
 &  & -{\cal I}(E_{c};w)\frac{A_{2}}{w_{1}}+\frac{1}{E_{c}}\frac{A_{3}}{ww_{0}}\,,
\end{eqnarray}
\begin{eqnarray}
W_{2}^{(0)}(w_{3}) & = & {\cal G}\left(E_{c};w_{3}\right)\frac{A_{0}}{w_{3}^{2}}-\frac{1}{E_{c}}\frac{A_{0}}{w_{3}}\,,\\
W_{4}^{(0)}(w_{0},w_{3}) & = & -{\cal G}(E_{c};w_{3})\frac{w_{3}A_{2}+w_{2}A_{3}}{w_{2}^{2}w_{3}^{2}}+{\cal G}(E_{c};w_{0})\frac{(w_{0}+w_{2})A_{2}+w_{2}A_{3}}{w_{0}^{2}w_{2}^{2}}\nonumber \\
 &  & -{\cal H}(E_{c};w_{0})\frac{A_{2}}{w_{0}w_{2}}+\frac{1}{E_{c}}\frac{A_{3}}{w_{0}w_{3}}\,,\\
W_{6}^{(0)}(w,w_{3}) & = & -{\cal G}(E_{c};w_{3})\frac{wA_{0}}{w_{3}^{2}(w^{2}-w_{3}^{2})}+{\cal G}(E_{c};w)\frac{w_{3}A_{0}}{w^{2}(w^{2}-w_{3}^{2})}+\frac{1}{E_{c}}\frac{A_{0}}{ww_{3}}\,,\label{eq:s6}
\end{eqnarray}
and 
\begin{eqnarray}
 &  & W_{8}^{(0)}(w,w_{0},w_{3})\nonumber \\
 & = & {\cal G}(E_{c};w_{3})\left[\frac{A_{2}}{(w-w_{3})w_{2}^{2}w_{3}}+\frac{w^{2}w_{2}+w_{3}^{3}+ww_{3}(-3w_{0}+2w_{3})}{(w-w_{3})^{3}w_{2}^{2}w_{3}^{2}}A_{3}\right]\nonumber \\
 & + & {\cal G}(E_{c};w_{0})\left[-\frac{w_{0}w_{1}+w_{1}w_{2}-w_{0}w_{2}}{w_{0}^{2}w_{1}^{2}w_{2}^{2}}A_{2}-\frac{w_{1}w_{2}-w_{0}^{2}-w_{0}w_{2}}{w_{1}^{2}w_{0}^{2}w_{2}^{2}}A_{3}\right]\nonumber \\
 & + & {\cal G}(E_{c};w)\left[-\frac{1}{ww_{1}^{2}(w-w_{3})}A_{2}-\frac{5w^{2}+w_{3}(w_{0}+w_{3})-w(3w_{0}+4w_{3})}{ww_{1}^{2}(w-w_{3})^{3}}A_{3}\right]\nonumber \\
 & + & {\cal H}(E_{c};w_{0})\left(\frac{A_{2}}{w_{0}w_{1}w_{2}}-\frac{A_{3}}{w_{1}^{2}w_{2}}\right)+{\cal H}(E_{c};w)\frac{4w^{2}-3ww_{0}-2ww_{3}+w_{0}w_{3}}{ww_{1}^{2}(w-w_{3})^{2}}A_{3}\nonumber \\
 & + & {\cal I}(E_{c};w)\frac{A_{3}}{w_{1}(w-w_{3})}-\frac{1}{E_{c}}\frac{A_{3}}{ww_{0}w_{3}}\,.\label{eq:s8}
\end{eqnarray}
\end{subequations}
Here we have used $w_{2}=w_{0}-w_{3}$, $w_{1}=w-w_{0}$, and $A_i$
defined in Eq.~(\ref{eq:defA}).

\subsection{$W_{j}^{(2)}$}

The $\Delta^{2}$ terms are given as \begin{subequations} 
\begin{eqnarray}
W_{1}^{(2)} & = & \frac{1}{E_{c}^{3}}A_{0}\,,\\
W_{3}^{(2)}(w_{0}) & = & \left[-\frac{1}{E_{c}^{3}}\frac{1}{2w_{0}}-\frac{1}{E_{c}}\frac{2}{w_{0}^{3}}+{\cal H}_{E_{c}}(w_{0})\frac{2}{w_{0}^{3}}\right](A_{1}+A_{2})\,,\\
W_{5}^{(2)}(w) & = & -\frac{1}{E_{c}^{3}}\frac{A_{0}}{w}-{\cal I}(E_{c};w)\frac{2A_{0}+2A_{1}}{w^{2}}+\frac{1}{E_{c}}\frac{4A_{1}}{w^{3}}-{\cal H}(E_{c};w)\frac{4A_{1}}{w^{3}}\,,\\
W_{7}^{(2)}(w,w_{0}) & = & \frac{1}{E_{c}^{3}}\frac{A_{1}+A_{2}}{2ww_{0}}+\frac{1}{E_{c}}\left[\frac{2(w^{2}-w_{0}^{2})(A_{1}+A_{3})}{w^{3}w_{0}^{3}}+\frac{2(w^{3}-w_{0}^{3})(A_{2}-A_{3})}{w_{1}w^{3}w_{0}^{3}}\right]
\nonumber \\
 &  & -{\cal I}(E_{c};w)\frac{2(A_{1}+A_{3})}{w^{2}w_{1}}+{\cal H}(E_{c};w)\frac{(-4w+2w_{0})(A_{1}+A_{3})+2w_{1}(A_{2}-A_{3})}{w^{3}w_{1}^{2}}\nonumber \\
 &  & +{\cal H}(E_{c};w_{0})\frac{-2w_{1}(A_{1}+A_{2})+2w_{0}(A_{1}+A_{3})}{w_{1}^{2}w_{0}^{3}}\,,
\end{eqnarray}
and 
\begin{eqnarray}
W_{2}^{(2)}(w_{3}) & = & \left\{ \frac{1}{E_{c}^{3}}\frac{1}{3w_{3}}+\frac{1}{E_{c}}\frac{4}{w_{3}^{3}}-{\cal G}(E_{c};w_{3})\frac{4}{w_{3}^{3}}\right\} A_{1}\,,\\
W_{4}^{(2)}(w_{0},w_{3}) & = & \frac{1}{E_{c}^{3}}\frac{-A_{1}+A_{2}-2A_{3}}{6w_{0}w_{3}}\nonumber \\
 &  & +\frac{1}{E_{c}}\left[-\frac{2(w_{0}^{2}-w_{3}^{2})(A_{1}+A_{3})}{w_{0}^{3}w_{3}^{3}}+\frac{2(w_{0}^{3}-w_{3}^{3})(A_{2}-A_{3})}{w_{2}w_{0}^{3}w_{3}^{3}}\right]\nonumber \\
 &  & -{\cal H}(E_{c};w_{0})\frac{A_{1}+A_{3}}{w_{0}w_{2}}\nonumber \\
 &  & +{\cal G}(E_{c};w_{0})\frac{2w_{0}(A_{1}+A_{3})+2w_{2}(A_{1}-A_{0})}{w_{0}^{4}w_{2}^{2}}\nonumber \\
 &  & +{\cal G}(E_{c};w_{3})\frac{(2w_{0}-4w_{3})(A_{1}+A_{3})-2w_{2}(A_{2}-A_{3})}{w_{2}^{2}w_{3}^{4}}\,,\\
W_{6}^{(2)}(w,w_{3}) & = & -\frac{1}{E_{c}^{3}}\frac{A_{1}}{3ww_{3}}+\frac{1}{E_{c}}\left[-\frac{4(w^{2}+w_{3}^{2})A_{1}}{w^{3}w_{3}^{3}}+\frac{4(A_{2}-A_{3})}{w^{2}w_{3}^{2}}\right]\nonumber \\
 &  & +{\cal G}(E_{c};w)\frac{-4w_{3}A_{1}+4w(A_{2}-A_{3})}{w^{4}(w^{2}-w_{3}^{2})}\nonumber \\
 &  & +{\cal G}(E_{c};w_{3})\frac{4wA_{1}-4w_{3}(A_{2}-A_{3})}{w_{3}^{4}(w^{2}-w_{3}^{2})}\,,
\end{eqnarray}
and 
\begin{eqnarray}
 &  & W_{8}^{(2)}(w,w_{0},w_{3})\nonumber \\
 & = & \frac{1}{E_{c}^{3}}\frac{A_{1}-A_{2}+2A_{3}}{6ww_{0}w_{3}}+\frac{1}{E_{c}}\left\{ \frac{2[w_{0}^{2}w_{3}^{2}+w^{2}(w_{0}^{2}-w_{3}^{2})](A_{1}-A_{2}+2A_{3})}{w^{3}w_{0}^{3}w_{3}^{3}}\right.\nonumber \\
 &  & \left.+\frac{2[w_{0}(2w_{0}+w_{3})+w(w_{0}+2w_{3})](-A_{2}+A_{3})}{w^{2}w_{0}^{3}w_{3}^{2}}\right\} \nonumber \\
 &  & +{\cal I}(E_{c};w)\frac{2(A_{1}+A_{2})}{w^{2}w_{1}(w-w_{3})}\nonumber \\
 &  & +{\cal H}(E_{c};w)\frac{2(6ww_{1}+ww_{2}-3w_{1}w_{3})(A_{1}+A_{2})}{w^{3}w_{1}^{2}(w-w_{3})^{2}}\nonumber \\
 &  & -{\cal G}(E_{c};w)\left[\frac{2(6ww_{1}+ww_{2}-3w_{1}w_{3})(A_{1}+A_{2})}{w^{3}w_{1}^{2}(w-w_{3})^{3}}-\frac{2(3w-2w_{0})}{w^{4}w_{1}^{2}(w-w_{3})}(A_{0}-A_{1})\right]\nonumber \\
 &  & -{\cal H}(E_{c};w_{0})\frac{2w_{1}(A_{1}+A_{3})-2w_{0}(A_{1}+A_{2})}{w_{1}^{2}w_{0}^{3}w_{2}}\nonumber \\
 &  & +{\cal G}(E_{c};w_{0})\frac{2ww_{0}(A_{0}-2A_{1}-A_{3})+2(3w_{0}^{2}+ww_{3}-2w_{0}w_{3})(A_{1}-A_{0})}{w_1^2w_0^4w_2^2}\nonumber \\
 &  & +{\cal G}(E_{c};w_{3})\left[\frac{2w(3ww_{0}-4ww_{3}-5w_{0}w_{3}+6w_{3}^{2})(-A_{1}+A_{2}-2A_{3})}{3(w-w_{3})^{3}w_{2}^{2}w_{3}^{4}}\right.\nonumber \\
 &  & \left.+\frac{2(w-3w_{3})(w+2w_{0}-3w_{3})(A_{1}-A_{0})}{3(w-w_{3})^{3}w_{2}^{2}w_{3}^{3}}\right]\,.
\end{eqnarray}
\end{subequations}

\subsection{$W_{j}^{(4)}$}

All terms proportional to $\Delta^{4}$ can be written as $W_{j}^{(4)}={\cal W}_{j}^{(4)}A_{0}$,
with all quantity ${\cal W}_{j}^{(4)}$ giving by \begin{subequations}
\begin{eqnarray}
{\cal W}_{1}^{(4)} & = & -\frac{1}{E_{c}^{5}}\frac{3}{2}\,,\\
{\cal W}_{3}^{(4)}(w_{0}) & = & -\frac{1}{E_{c}^{5}}\frac{1}{2w_{0}}-\frac{1}{E_{c}^{3}}\frac{2}{w_{0}^{3}}-\frac{1}{E_{c}}\frac{8}{w_{0}^{5}}+{\cal H}(E_{c};w_{0})\frac{8}{w_{0}^{5}}\,,\\
{\cal W}_{5}^{(4)}(w) & = & \frac{1}{E_{c}^{5}}\frac{3}{2w}+\frac{1}{E_{c}^{3}}\frac{2}{w^{3}}-\frac{1}{E_{c}}\frac{8}{w^{5}}+{\cal I}(E_{c};w)\frac{8}{w^{4}}+{\cal H}(E_{c};w)\frac{8}{w^{5}}\,,\\
{\cal W}_{7}^{(4)}(w,w_{0}) & = & \frac{1}{E_{c}^{5}}\frac{1}{2ww_{0}}+\frac{1}{E_{c}^{3}}\frac{2(w^{2}-w_{0}^{2})}{w^{3}w_{0}^{3}}+\frac{1}{E_{c}}\frac{8(w^{4}-w^{2}w_{0}^{2}-2ww_{0}^{3}-3w_{0}^{4})}{w^{5}w_{0}^{5}}\nonumber \\
 &  & -{\cal I}(E_{c};w)\frac{8}{w^{4}w_{1}}-{\cal H}(E_{c};w)\frac{8(w+3w_{1})}{w^{5}w_{1}^{2}}-{\cal H}(E_{c};w_{0})\frac{8(w-2w_{0})}{w_{1}^{2}w_{0}^{5}}\,.
\end{eqnarray}
\begin{eqnarray}
{\cal W}_{2}^{(4)}(w_{3}) & = & -\frac{1}{E_{c}^{5}}\frac{1}{10w_{3}}-\frac{1}{E_{c}^{3}}\frac{2}{3w_{3}^{3}}-\frac{1}{E_{c}}\frac{8}{w_{3}^{5}}+{\cal G}(E_{c};w_{3})\frac{8}{w_{3}^{6}}\,,\\
{\cal W}_{4}^{(4)}(w_{0},w_{3}) & = & -\frac{1}{E_{c}^{5}}\frac{3}{10w_{0}w_{3}}-\frac{1}{E_{c}^{3}}\frac{2(3w_{0}^{2}+2w_{0}w_{3}+w_{3}^{2})}{3w_{0}^{3}w_{3}^{3}}-\frac{1}{E_{c}}\frac{8(3w_{0}^{4}+2w_{0}^{3}w_{3}+w_{0}^{2}w_{3}^{2}-w_{3}^{4})}{w_{0}^{5}w_{3}^{5}}\nonumber \\
 &  & -{\cal H}(E_{c};w_{0})\frac{8}{w_{0}^{5}w_{2}}+{\cal G}(E_{c};w_{0})\frac{8(3w_{0}-2w_{3})}{w^{6}w_{2}^{2}}+{\cal G}(E_{c};w_{3})\frac{8(3w_{0}-4w_{3})}{w_{2}^{2}w_{3}^{6}}\,,\\
{\cal W}_{6}^{(4)}(w,w_{3}) & = & \frac{1}{E_{c}^{5}}\frac{1}{10ww_{3}}+\frac{1}{E_{c}^{3}}\frac{2(w^{2}+w_{3}^{2})}{3w^{3}w_{3}^{3}}+\frac{1}{E_{c}}\frac{8(w^{4}+w^{2}w_{3}^{2}+w_{3}^{4})}{w^{5}w_{3}^{3}}\nonumber \\
 &  & +{\cal G}(E_{c};w)\frac{8w_{3}}{w^{6}(w^{2}-w_{3}^{2})}-{\cal G}(E_{c};w_{3})\frac{8w}{w_{3}^{6}(w^{2}-w_{3}^{2})}\,,
\end{eqnarray}
and 
\begin{eqnarray}
{\cal W}_{8}^{(4)}(w,w_{0},w_{3}) & = & \frac{1}{E_{c}^{5}}\frac{3}{10ww_{0}w_{3}}+\frac{1}{E_{c}^{3}}\frac{2\left[w^{2}\left(3w_{0}^{2}+2w_{0}w_{3}+w_{3}^{2}\right)-w_{0}^{2}w_{3}^{2}\right]}{3w^{3}w_{0}^{3}w_{3}^{3}}\nonumber \\
 &  & +\frac{1}{E_{c}}\frac{8\left(\left(3w_{0}^{4}+2w_{3}w_{0}^{3}+w_{3}^{2}w_{0}^{2}-w_{3}^{4}\right)w^{4}+\left(w_{3}^{2}-w_{0}^{2}\right)w^{2}w_{0}^{2}w_{3}^{2}+w_{0}^{3}w_{3}^{4}(2w+3w_{0})\right)}{w^{5}w_{0}^{5}w_{3}^{5}}\nonumber \\
 &  & +{\cal I}(E_{c};w)\frac{8}{w^{4}w_{1}(w-w_{3})}+{\cal H}(E_{c};w)\frac{8(8w^{2}-7ww_{0}-6ww_{3}+5w_{0}w_{3})}{w^{5}w_{1}^{2}(w-w_{3})^{2}}\nonumber \\
 &  & -{\cal G}(E_{c};w)\frac{8[18w^{3}-8w_{0}w_{3}^{2}+ww_{3}(21w_{0}+10w_{3})-w^2(15w_0+26w_3)]}{w^{6}w_{1}^{2}(w-w_{3})^{3}}\nonumber \\
 &  & +{\cal H}(E_{c};w_{0})\frac{8\left(w-2w_{0}\right)}{\left(w-w_{0}\right){}^{2}w_{0}^{5}\left(w_{0}-w_{3}\right)}-{\cal G}(E_{c};w_{0})\frac{8\left(w-2w_{0}\right)(3w_0-2w_3)}{\left(w-w_{0}\right){}^{2}w_{0}^{6}\left(w_{0}-w_{3}\right)^2}\nonumber \\
 &  & +{\cal G}(E_{c};w_{3})\frac{8\left[\left(4w_{3}-3w_{0}\right)w^{2}+3\left(3w_{0}-4w_{3}\right)w_{3}w+2w_{3}^{2}\left(5w_{3}-4w_{0}\right)\right]}{\left(w-w_{3}\right){}^{3}\left(w_{0}-w_{3}\right){}^{2}w_{3}^{6}}\,.\quad
\end{eqnarray}
\end{subequations}

\bibliographystyle{apsrev4-1}

%

\end{document}